\documentclass[review]{elsarticle}

\journal{Joule}

\usepackage[superscript,biblabel]{cite}
\usepackage{tabularx}

\usepackage{mathtools}

\begin{document}

\begin{frontmatter}

\title{30.000 ways to reach 55\% decarbonization of the European electricity sector
               }

\author{Tim T. Pedersen \fnref{MPE,iclimate} \corref{mycorrespondingauthor}}
\fntext[MPE]{Department of Mechanical and Production Engineering, Aarhus University, Aarhus, Denmark}
\cortext[mycorrespondingauthor]{Corresponding author, Email: ttp@mpe.au.dk}

\author{Mikael Skou Andersen \fnref{ENV, iclimate}}
\fntext[ENV]{Department of Environmental Science - Environmental social science and geography, Aarhus University, Foulum, Denmark}

\author{Marta Victoria \fnref{MPE,iclimate}}

\author{Gorm B. Andresen \fnref{MPE,iclimate} }
\fntext[iclimate]{ICLIMATE Interdisciplinary Centre for Climate Change, Aarhus University, Aarhus, Denmark}


\begin{graphicalabstract}
\includegraphics[width=1\textwidth]{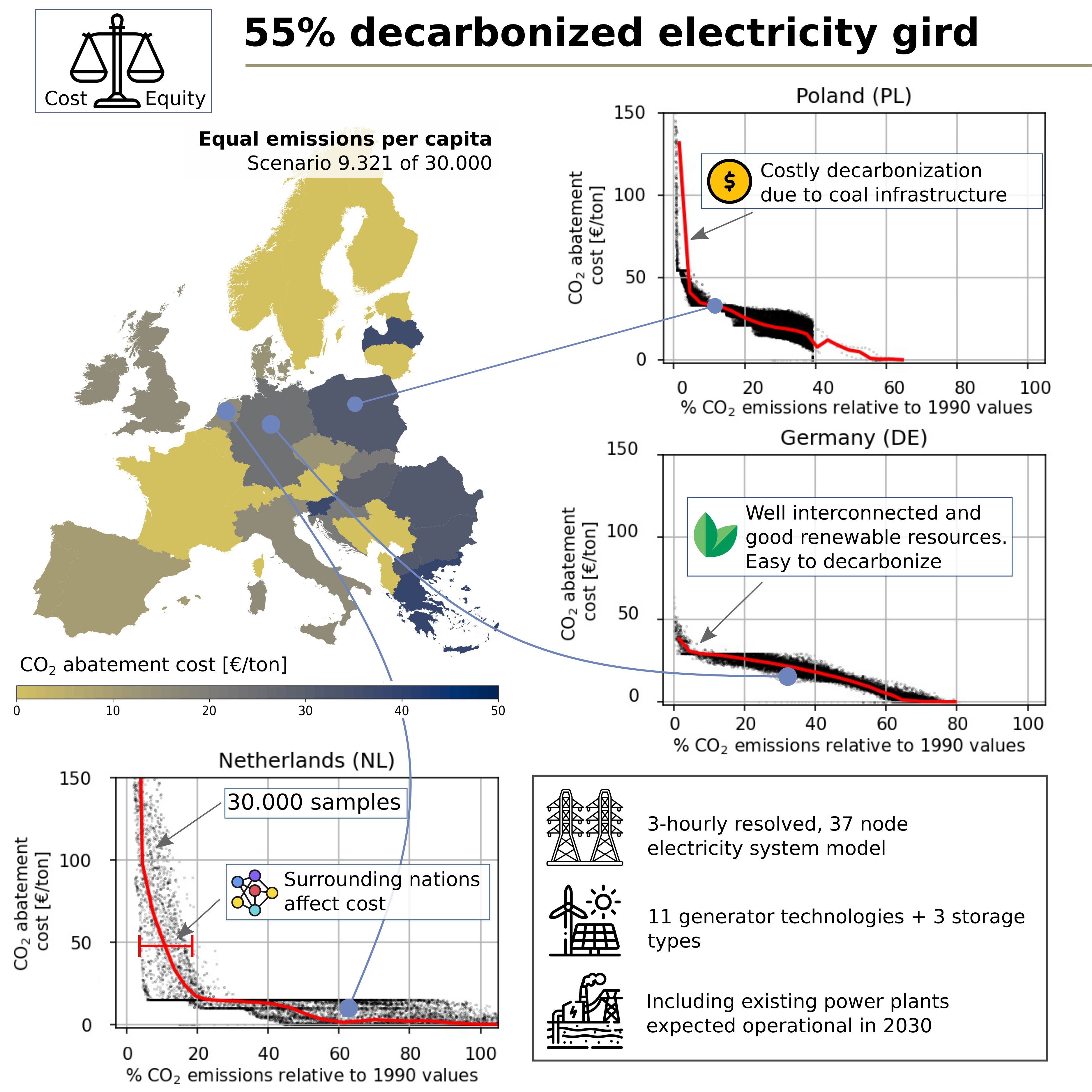}
\end{graphicalabstract}

\begin{abstract} 
Climate change mitigation is a global challenge that, however, needs to be resolved by national-level authorities, resembling a “tragedy of the commons”. This paradox is reflected at European scale, as climate commitments are made by the EU collectively, but implementation is the responsibility of individual Member States.  Here, we investigate 30.000 near-optimal effort-sharing scenarios where the European electricity sector is decarbonized by at least 55\% relative to 1990, in line with 2030 ambitions. Using a highly detailed brownfield electricity system optimization model, the optimal electricity system is simulated for a suite of effort-sharing scenarios. Results reveal large inequalities in the efforts required to decarbonize national electricity sectors, with some countries facing cost-optimal pathways to reach 55\% emission reductions, while others are confronted with relatively high abatement costs. Specifically, we find that several countries with modest or low levels of GDP per capita will experience high abatement costs, and when passed over into electricity prices this may lead to increased energy poverty in certain parts of Europe.
\end{abstract}

\begin{highlights}
   \item Achieving a just transition under homogeneous CO$_2$ price requires additional regulation.
   \item Nations with high CO$_2$ abatement costs are likely to have high electricity prices
   \item Larges CO$_2$ abatement costs observed in nations with low GDP
   \item Countries where emissions are hard to reduce are likely to have high emission intensities and power prices. 
   \item The realized energy system is likely to be 5\% more expensive than the theoretical optimum. 
\end{highlights}

\end{frontmatter}


\section{Introduction}
One of the most crucial aspects of climate change mitigation is how to overcome “the tragedy of the commons” \cite{nordhaus2019}. Climate change is a global problem that cannot be managed without deliberate action from all individual countries on Earth.
Agreements to reduce emissions have historically been made as multilateral agreements, e.g., global agreements such as the Kyoto Protocol and the Paris agreement, or lately the European Green Deal \cite{green_deal}. The implementation of regulations to reduce emissions is, however, to be done at a national level. As strict national targets for emissions reductions are likely to entail a loss in welfare in the short term, the inclination of most countries so far has been to opt for the least ambitious emission target that satisfies their commitment to global agreements \cite{van2020implications}. This can be observed today where the sum of nationally determined contributions, is far from sufficient to meet the targets of the Paris agreement \cite{NDC_report}. In other words, nations are maximizing their national welfare, but in doing so the common resource, our atmosphere, is exhausted prematurely. This behavior will lead to a decrease in global welfare in the long run, as climate change eventually will entail much higher costs, than what is gained from presently uncurbed emissions \cite{tol2009economic}.\\

Harding argues that there exist no technical solutions that can be found to solve such a problem of the commons, other than by dividing up the common resource \cite{hardin1968tragedy}. The finite resource - that is our atmosphere's CO$_2$ carrying capability - can be divided through the implementation of a cap-and-trade policy. 
Setting a quota on emissions that in turn will allow emission rights to be traded in a market with a carbon price should lead to efficient decarbonization of our society \cite{nordhaus2019}, but at what cost? Bauer et. al. \cite{Bauer_2020} find that huge international financial transfers are in fact required to ensure justice of the transition, when governed by a cap-and-trade system. This significant but unresolved equity issue is replicated at regional level in Europe, where climate commitments are made by the EU, but implementation and governance rest with individual countries. In the European Union, there is now a commitment to aim for a just transition as stated in the Green Deal \cite{green_deal}. Financial support will be provided by the Just Transition Mechanism to those jurisdictions where the socio-economic impact of the transition will be highest \cite{just_transition_mechanism}. The quality of a just transition is, however, not easily measured and has in recent years become a topic of much debate \cite{jenkins_2016,maguire_2021}.\\

The European power sector is in the early phase of a major transformation from a fossil-fuel-based system to relying mainly on renewable and low-carbon resources \cite{victoria_2020}. However, efficient efforts to reduce emissions differ between countries that have different starting points, public preferences, geographical constraints, etc. Some countries have already installed high shares of carbon-neutral energy sources, while other countries are deeply reliant on fossil fuels \cite{eurostat_2019}. A historical example of the diversity among national strategies toward transforming their energy supply was seen in the response to the oil crisis in the 70s. Here vastly different strategies to reduce reliance on oil imports were employed. While Brazil took advantage of the low demand for locally produced sugarcane and started large-scale production of bio-ethanol, France invested heavily in nuclear infrastructure \cite{solomon2011coming}. The historical path-dependency in national energy supply not only sets the starting point for the green transition but can also breed an ‘energy culture’ with a preference for specific technologies. Denmark, for instance, became a pioneer in wind power thanks to R\&D efforts that began as early as the 1880s with Paul La Cour’s novel blade design \cite{dyrhauge2017denmark}.

\begin{figure}[!htb]
   \includegraphics[width=1\textwidth,trim={0cm 0cm 0cm 0cm},clip]{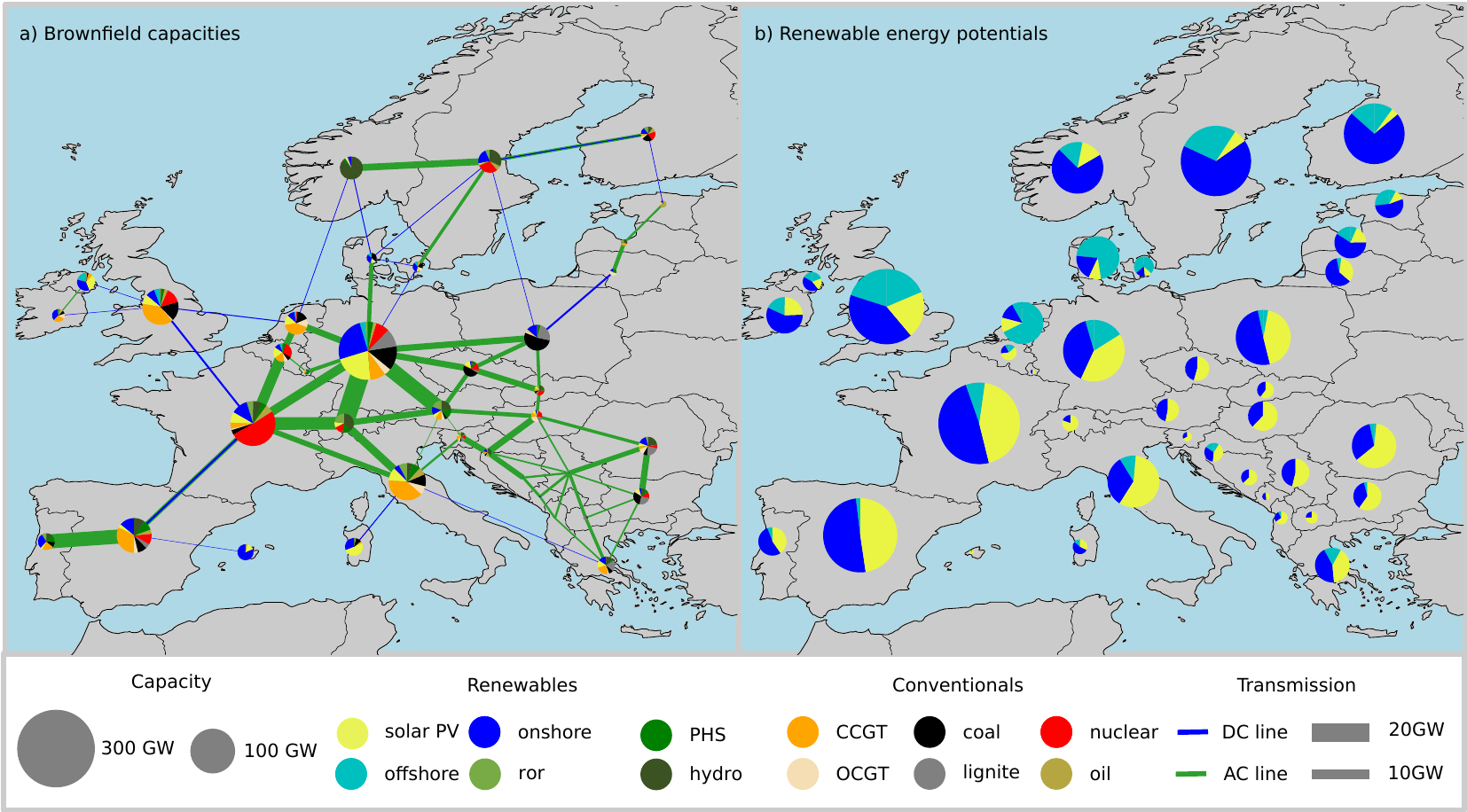}
   \caption{\textbf{Existing generator capacity and renewable potentials -} a) Currently installed technology capacity that is expected to be in operation in 2030. Capacities represent maximum electricity generation. b) Effective renewable energy potentials, calculated as the maximum geographical potential times the local capacity factor for the given technology. Wind turbine capacity is specified as offshore and onshore. Hydro-power is separated as either run-of-river (ROR), pumped-hydro-storage (PHS), or hydro. Two types of gas turbines are included; closed-cycle-gas-turbines (CCGT) and open-cycle-gas-turbines (OCGT). Transmission capacities are indicated either as high voltage AC or DC transmission lines. The majority of countries have both brownfield and renewable potentials to cover several times their annual electricity demand. }
   \label{fig:starting_point}
\end{figure}

The green transformation of Europe's domestic energy supply is challenged by several factors of technical, economic, and political nature. The access to renewable resources \cite{levenda_2021}, power plants currently in operation, and availability of international transmissions connections \cite{golubchikov_2020}. Figure \ref{fig:starting_point} a) shows the power and transmission capacity installed today and expected to remain in operation by 2030. It is clear, that the preconditions for rapid decarbonization are very diverse, with some countries relying heavily on coal and oil, while others have a large share of renewable power generation. Figure \ref{fig:starting_point} b) shows the potentials for renewables per Member State. Renewable potentials are calculated as the geographical potential for renewable capacity multiplied by the average national capacity factor for the given renewable resource. It is evident that the renewable potentials are unrelated to the nations population size and local demand. 
Given the diverse preconditions for a green transition in the European electricity supply, varied costs associated with the transition must be expected. Ensuring equity in the transition requires the identification of nations where the green transition comes at a higher cost and nations where only minor political incentives are required. \\

A uniform CO$_2$ price is considered the most efficient way to achieve emissions reductions \cite{bohringer2009eu}. However, with a uniform CO$_2$ price, the fairness of the transformation cannot be ensured. The cost-efficient solution obtained with a uniform CO$_2$ price is likely to favor early decarbonization in regions where renewable resources are favorable, resulting in a skewed transformation \cite{grunewald2017renewable}.
If equal effort sharing is to be ensured under a uniform CO$_2$ price, financial transfers are required as the cost of decarbonizing the energy supply varies between nations \cite{Bauer_2020}. 
Agreeing upon what equal sharing of effort looks like is inherently difficult as it is not a question about costs but rather one of ethics \cite{sovacool2015energy}. Zhou and Wang \cite{Zhou_2016} identify a range of effort-sharing schemes, based on different principles, such as sovereignty, egalitarianism, efficiency, horizontal equity, vertical equity, and polluter pays. Other studies combine these principles to create more complex effort-sharing schemes, such as the Model of Climate Justice per capita \cite{alcaraz_2018}, where historical emissions along with population growth are considered. 
Markowitz notes that we are ill-equipped to decide, given the complexity of the problem, and our own complicity in causing it \cite{markowitz2012climate}. Still, Jenkins et al. identifies three core tenets of energy justice called: distributional, recognition, and procedural justice \cite{jenkins_2016}. Distributional justice recognizes the unequal distribution of environmental benefits and provides the rationale for this research. Recognition justice states that individuals must be fairly represented and have equal rights, whereas procedural justice concerns access to the decision-making process. In this work, the difficulties of addressing distributional justice in the transformation of the European electricity supply under a global CO$_2$ price will be addressed. \\

Several studies exist where the equity efficiency trade-off occurring in the transition of our energy supply has been conducted. 
Bauer et al. \cite{Bauer_2020} studied the trade-off to be made between economic efficiency and sovereignty in a global context using a multi-objective approach combined with an integrated assessment model. Their findings indicate a highly non-linear trade-off between efficiency and sovereignty, where the intermediate scenarios can secure higher total benefits. 
Van den Berg et al. \cite{van2020implications} studied the implications of employing several effort-sharing approaches in a global context, finding a large discrepancy between the cost-optimal solution and the effort-sharing principles. 
Schwenk-Nebbe and co-authors \cite{Schwenk_2020}, consider three different effort-sharing principles for establishing national CO$_2$ reduction targets in a European context. 
In the Efficiency solution, they find that the remaining emissions will be concentrated in a small number of countries where the costs of decarbonization are highest. Allocating emissions after a Sovereignty or Grandfathering principle will distribute the remaining emissions more evenly while implying a spread in-unit abatement costs and higher total system cost. 
Sasse et al. \cite{sasse2019distributional} study the trade-off between regional equity and efficiency in the case of Switzerland's energy supply. Their findings show how renewable energy production will be concentrated in regions with favorable renewable resources if no equity measures are employed. 
A similar study for the German electricity sector has been conducted drawing similar conclusions \cite{drechsler2017efficient}.

The studies conducted seek to address the problem of the equity-efficiency trade-off by imposing varying effort-sharing principles. The underlying problem is, however, that the cost of abating emissions is not equally distributed. Thus there is a knowledge gap in studies investigating the inequalities in abatement costs. To our knowledge, there exists no detailed study on the diversity of transition costs in a European context. 

The aim of this paper is to answer the question of, how the cost of abating CO$_2$ emissions in the electricity sector are distributed across Europe. 
This is achieved by using a detailed model of the European electricity sector to investigate how the CO$_2$ abatement cost in each specific country is related to its corresponding national emission target as well as to a wide combination of targets for the other countries in the union. By analyzing how the costs associated with the decarbonization of the electricity sector are distributed among European countries, and what effect changing national emission targets have on the national energy supplies, the magnitude of policy push required to obtain equitable effort sharing is obtained. 

The novelty of this work lies in the combination of an energy system optimization model and effective Markov Chain Monte Carlo (MCMC) methods to study all possible configurations of national emission targets in a European context. Where previous studies have used scenario-based modeling or multi-objective optimization to study the implications of a small range of effort-sharing schemes, the method applied in this work is capable of addressing the underlying problem structure by investigating all possible outcomes. Furthermore, detailed information about each configuration of national reduction targets can be obtained, as the power system optimization model is solved for each allocation scheme.

\section{Near-optimal solutions in PyPSA-Eur-Sec}
In this research, the Adaptive Metropolis-Hastings (AMH) method \cite{haario_2001} has been employed in combination with the PyPSA-Eur-Sec \cite{PyPSA-EUR-SEC} techno-economic optimization model of the European electricity supply system. The goal is to identify and study all near-optimal configurations of national emission targets under which the EU can reach its climate targets as a whole. 
Focus is placed on near-optimal configurations, meaning national emission target configurations that do not increase the total system cost relative to the cost-optimal solution by more than 18\%. By identifying all near-optimal configurations of national emission targets, the national costs related to different levels of decarbonization can be identified. 
Four criteria are required for all the configurations as listed in Table \ref{tab:feasa}. Most importantly, a joint CO$_2$ reduction of at least 55\% must be achieved by the model countries in line with the EU’s 2030 Climate Target Plan \cite{EU_2030_Climate_target_plan}. Furthermore, the total system cost of implementing the reduction targets must not increase by more than 18\% relative to the cost-optimal allocation of targets. This constraint is based on the principles from Modeling to Generate Alternatives (MGA) where economically near-optimal model solutions are studied \cite{brill1982}. Applying MGA methods to energy system optimization models has gained a lot of attention recently \cite{neumann2021broad, lombardi2020policy}. MGA has, however, mainly been used to study technical flexibility among the near-optimal solutions \cite{neumann2021}. 

\begin{table*}[htb]
\centering
\caption{CO$_2$ configurations scheme feasibility criteria}
\label{tab:feasa}
\begin{tabularx}{.9\textwidth}{lX}
\hline
a) & The joint CO$_2$ reductions must be equal to or greater than 55\% relative to 1990.  \\
b) & Total system cost of the configuration of national reductions should not exceed the cost-optimal scenario with 18\%.  \\
c) & A technically feasible solution to the model exists.  \\
d) & National emissions must remain below the equivalent of supplying 150\% of energy demand with coal.
\end{tabularx}
\end{table*}

The energy system optimization model used is the PyPSA-Eur-Sec \cite{PyPSA-EUR-SEC} model spanning 33 ENTSO-E (European Network of Transmission System Operators for Electricity) member countries, i.e. the model includes EU-27 without Cyprus and Malta, instead including Norway, Switzerland, Serbia, Bosnia-Herzegovina, Albania, Montenegro, Macedonia and, United Kingdom. The model assumes long-term market equilibrium as well as perfect competition and foresight. A 2030 brownfield scenario is modeled, where all installed generator capacities as of 2019 that are expected to be in operation in 2030 are included. To cover energy demands, the model will install a new generation capacity, where it is economically optimal. The model uses a one node per synchronous zone setup, with the nodes connected by high voltage AC and DC lines. Using one year of energy demand and weather data resolved in 3-hour time steps, the model determines the cost-optimal dispatch, power flows, and investment in new generator capacity. Transmission line capacities included in the model are the currently installed capacities, plus the planned capacities from the Ten Year Network Development Plan (TYNDP 2018) \cite{tyndp_2018}. The energy-generating technologies included are hydro, onshore wind, offshore wind, solar PV, CCGT, OCGT, coal, lignite, nuclear, and oil. Furthermore, two storage technologies are included. These are hydrogen and battery storage. The technology parameters are listed in \ref{app:assumptions}. Brownfield capacities are shown in Figure \ref{fig:starting_point} and in Table \ref{tab:brownfield_cap}. The national CO$_2$ reduction targets provided by the AMH sampler are included as constraints in the model, limiting CO$_2$ emissions from energy generation in each of the modeled countries. Still, modeled countries are free to over-perform on the national CO$_2$ reduction target if it is economically favorable.

In a linear optimization problem - such as the one used here - the results obtained by imposing a constraint on, e.g., CO$_2$ emissions, could also be achieved by introducing a CO$_2$ price equivalent to the shadow price given by the constraint. Alternatively, the same result could also be obtained by subsidizing renewable technologies by the right amount or by imposing specific regulations on certain technologies. Thus, the scenarios considered in this work where national emission constraints are supplied could also be interpreted as scenarios with a combined policy push from different levels of subsidies, regulations, and CO$_2$ price. 

\section{Results}

\begin{figure}[htb]
               \centering
               \includegraphics[width=0.7\textwidth,trim={0 0 0 0},clip]{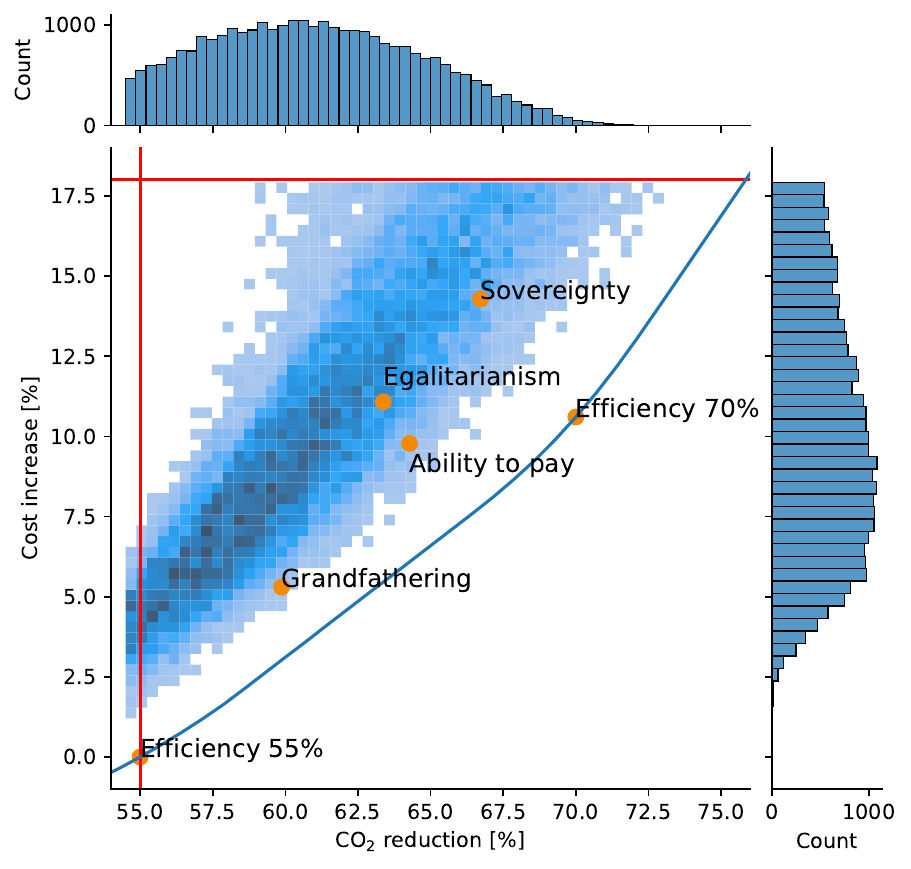}
               \caption{\textbf{Distribution of possible CO$_2$ reduction configurations -} Histogram showing the CO$_2$ reductions relative to 1990 and the associated costs of all feasible configurations of national reduction targets. The minimum required CO$_2$ reduction and maximum allowable cost increase is marked with red lines. The blue line marks the Pareto-optimal front of a dual objective optimization procedure using total system cost and joint CO$_2$ reduction as the two objective functions. The reduction target scenarios Efficiency 55\%, Efficiency 70\% Grandfathering, Ability-to-pay, Egalitarianism, and Sovereignty are marked with orange.} 
               \label{fig:co2red_cost}
\end{figure}

Applying the AMH method, a total of 30.000 near-optimal configurations of national emission targets were drawn randomly and unbiasedly. In Figure \ref{fig:co2red_cost}, the resulting realized CO$_2$ emissions from all configurations are shown, plotted against their total system costs. 
Five principle effort sharing schemes, inspired by \cite{Zhou_2016}, are shown as a reference. The principle schemes and their associated effort sharing rules are found in Table \ref{tab:scenarios}. The Efficiency scheme - representing a uniform CO$_2$ price - is implemented with two global emission reduction goals, namely 55\% and 70\% reductions. 
The reference scenario with the lowest total system cost is the Efficiency 55\% scenario. Further detail on the principle effort sharing schemes is found in the Experimental Procedures. 
In Figure \ref{fig:co2red_cost}, the Pareto front showing the cost versus CO$_2$ reduction trade-off is shown with a blue line. The Pareto optimal front was calculated by continuously decreasing the targeted joint CO$_2$ emissions using a uniform CO$_2$ price.

\begin{table*}[htb]
\caption{Strategies for CO$_2$ target configurations}
\label{tab:scenarios}
\begin{tabularx}{1\textwidth}{XXX}
Name           & Interpretation                                                       & Rule                                                          \\ \hline
Grandfathering & All nations have equal right to pollute                              & Distribute emissions proportionally to historical emissions   \\
Sovereignty     & All nations have equal right to pollute                              & Distribute emissions proportionally to energy demand    \\
Efficiency     & Maximize global welfare                                              & Distribute emissions to reduce total socio-economic costs \\
Egalitarianism & All citizens have equal right to pollute                               & Distribute emissions proportionally to population size      \\
Ability to pay & Nations with higher welfare should take on a larger part of the task & Distribute emissions inversely to GDP per capita    \\  \hline
\end{tabularx}
\end{table*}

In Figure \ref{fig:co2red_cost}, a gap between the Pareto optimal front and the samples can be observed. There is nothing preventing the sampler from identifying configurations on the Pareto optimal front, it is, however, very unlikely. This shows that the optimal solution is an extreme scenario that is very hard to obtain without extensive collaboration and agreement between all European countries. This is very unlikely, as countries have individual national targets and agendas. Therefore, we argue that solutions located in the dark blue regions of Figure \ref{fig:co2red_cost} can be considered significantly more probable outcomes because they can be realized with many different configurations of national emission targets.

All the principle effort sharing schemes, except Efficiency 55\%, are seen to provide a higher CO$_2$ reduction than required. This over-performance on emissions reduction is a result of several countries finding it cost-optimal to reduce emissions beyond their assigned national target. Further information on this behavior is available in the supplementary Figure \ref{fig:unused_co2}.
The Grandfathering and Ability to pay schemes are located relatively close to the Pareto-optimal front, whereas the Sovereignty and Egalitarianism are found to be further from it (see Figure \ref{fig:co2red_cost} and Figure \ref{fig:all55_scenarios}). The distance to the Pareto front can be interpreted as a measure of the efficiency of the scheme.

Considering the distribution of the joint CO$_2$ reduction, it is clear that the probability of achieving reductions close to the joint target (marked by the vertical red line) is more likely than overachieving. Moreover, the distribution of the system costs reveals that an increase in the total system costs of not less than 5\% relative to the cost-optimal scenario is almost unavoidable, since, to obtain the lowest possible total system costs, the burden of transitioning must be shared in a very exact way. It is, nevertheless, very unlikely that this will happen as countries have different national ambitions. Therefore costs higher than what is deemed optimal are to be expected. What Figure \ref{fig:co2red_cost} also shows, is that this cost increase with a 75\% probability will be above 4.6\% and it has a 50\% probability of being between 4.6\% and 12.2\%. \newline

\begin{figure}[htb]
    \includegraphics[width=1\textwidth,trim={1.2cm 0.6cm 2.5cm 1.2cm},clip]{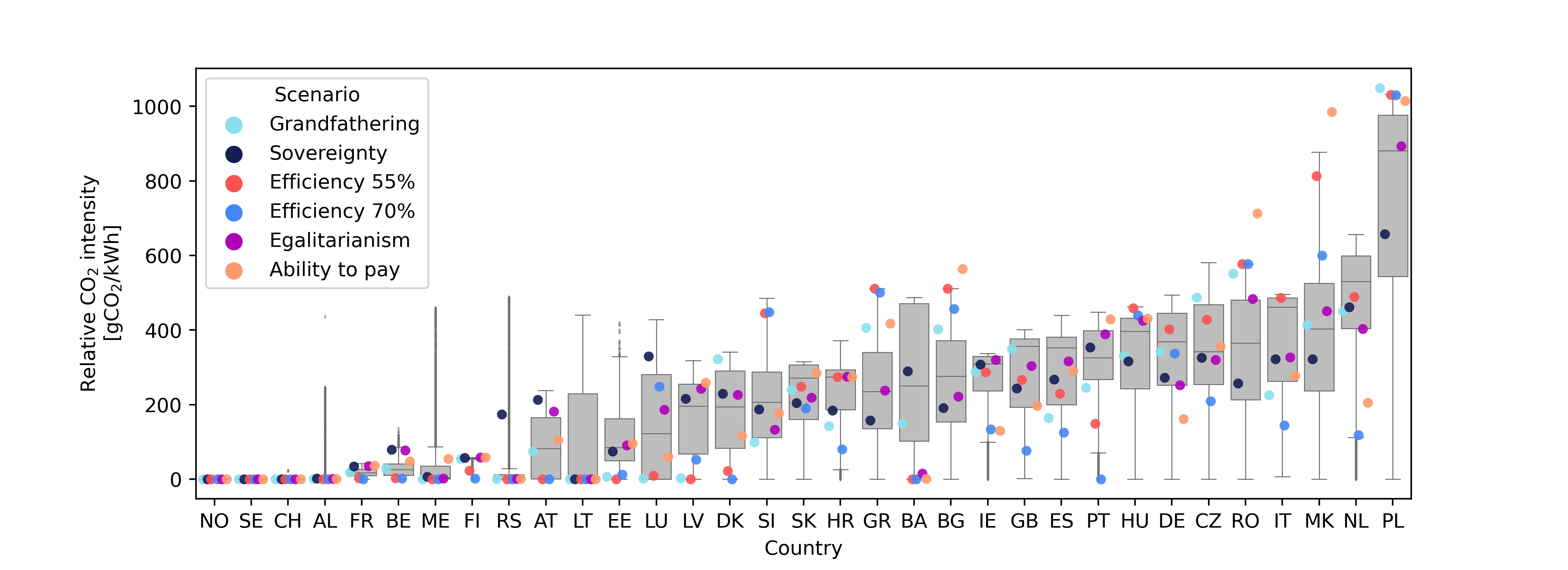}
   \caption{\textbf{National emission intensity -} National CO$_2$ emissions intensity for all modeled scenarios. Emission intensity is measured as emitted CO$_2$ pr MWh of produced electricity. The scenarios Grandfathering, Sovereignty, Efficiency, Egalitarianism, and Ability to pay is highlighted.}
   \label{fig:co2_mwh_box}
\end{figure}

Figure \ref{fig:co2_mwh_box} shows the probability range for CO$_2$ intensity (emissions per kWh) in each of the modeled countries according to the various configurations of national reduction targets. The average emission intensity for the EU-27 electricity supply was 230~gCO$_2$/kWh in 2020 according to EEA \cite{EEA_emissions}.
As seen in the figure, all countries have zero emissions in one or more configurations. Countries such as Norway and Sweden have zero emissions under all circumstances. This is not because they are allocated a demanding reduction target, but simply because it is cost-optimal to rely fully on renewable or nuclear energy in the electricity supply of these countries. On the other hand, countries such as Poland and North Macedonia tend to have large emissions intensity in most configurations (scenarios, where these countries have ambitious reduction targets, are very likely to be too expensive and thereby rejected). In 2020, the emission intensity for Poland's electricity supply was 700~gCO$_2$/kWh \cite{EEA_emissions}.
By analyzing the configuration of national reduction targets in the Efficiency approaches, it can be observed that higher than average shares of emissions are allocated to countries that at the outset have high emissions, and less than average shares of emissions are allocated to countries with low initial emissions. In other words, the Efficiency schemes favor assigning modest reduction targets to countries that have a hard time reducing emissions and cutting emissions drastically in countries where CO$_2$ reduction is easier. This intuitively reduces total system cost. When comparing the Efficiency scenario with 55\% and 70\% reductions, we can identify countries that are next in line to reduce emissions such as the Netherlands, Italy, and Portugal, as they all reduce emissions substantially in the 70\% scenario compared to the 55\% scenario. There are, however, also countries where the emission intensity is undisturbed even though joint emissions have been reduced. 
In the Sovereignty configuration, CO$_2$ reduction targets are assigned equally based on the national energy demand. Naturally, a much more even emissions intensity results from this approach. Poland and the Netherlands are, however, observed to have higher emission intensities than the other countries. 
In the Ability to pay approach, emissions are distributed inversely proportional to national GDP per capita. This redistribution of emissions with the Ability to pay approach is clearly discerned in Figure \ref{fig:co2_mwh_box}, where wealthy countries such as Germany and the Netherlands end up with low emissions while countries such as Romania, Macedonia, Poland, and Bulgaria feature higher emissions.\newline

\begin{figure}[htb]
               \includegraphics[width=0.5\textwidth,trim={3.2cm 3cm 1.1cm 2cm} ,clip] {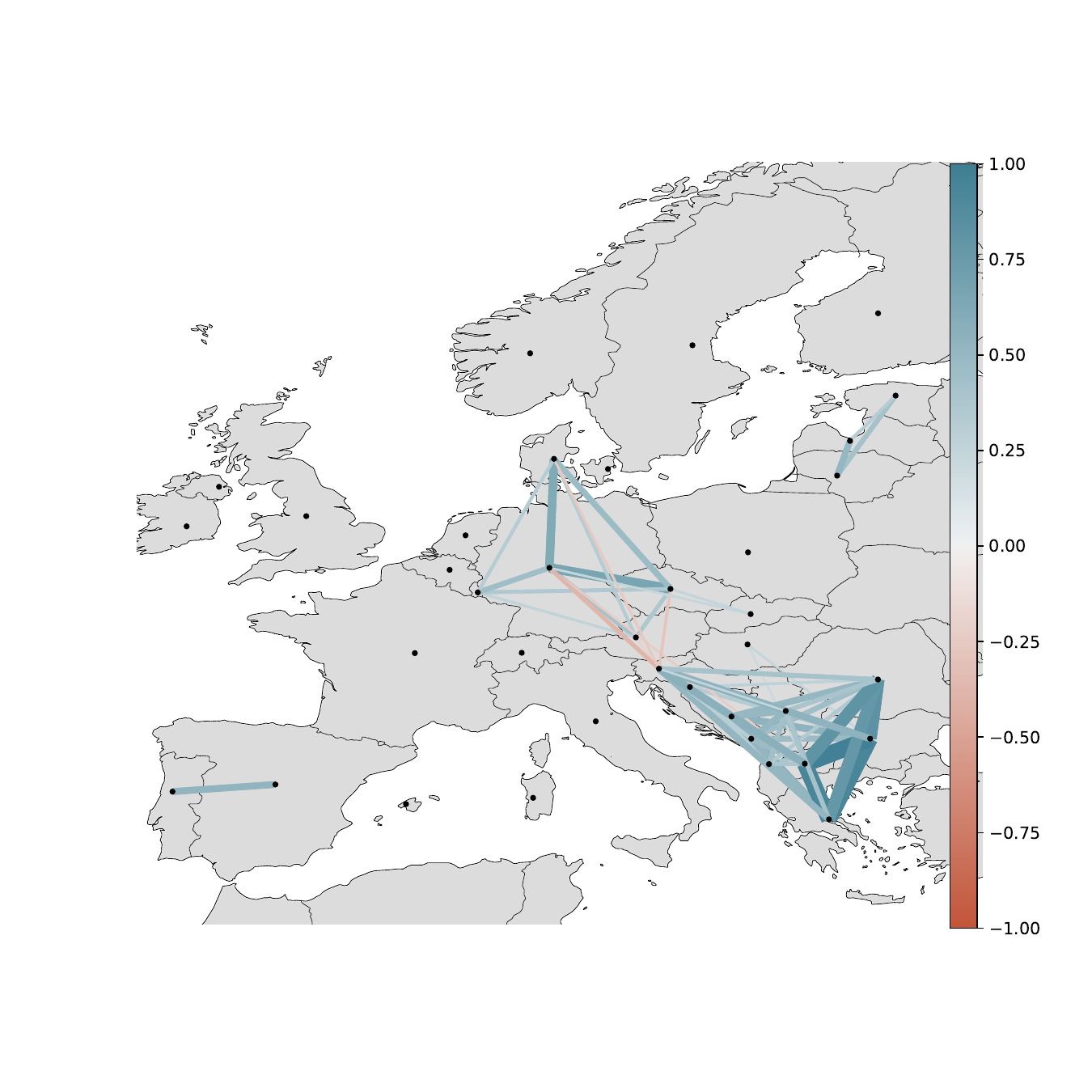}
               \includegraphics[width=0.5\textwidth,trim={1cm 1.5cm 1.5cm 1.5cm} ,clip]{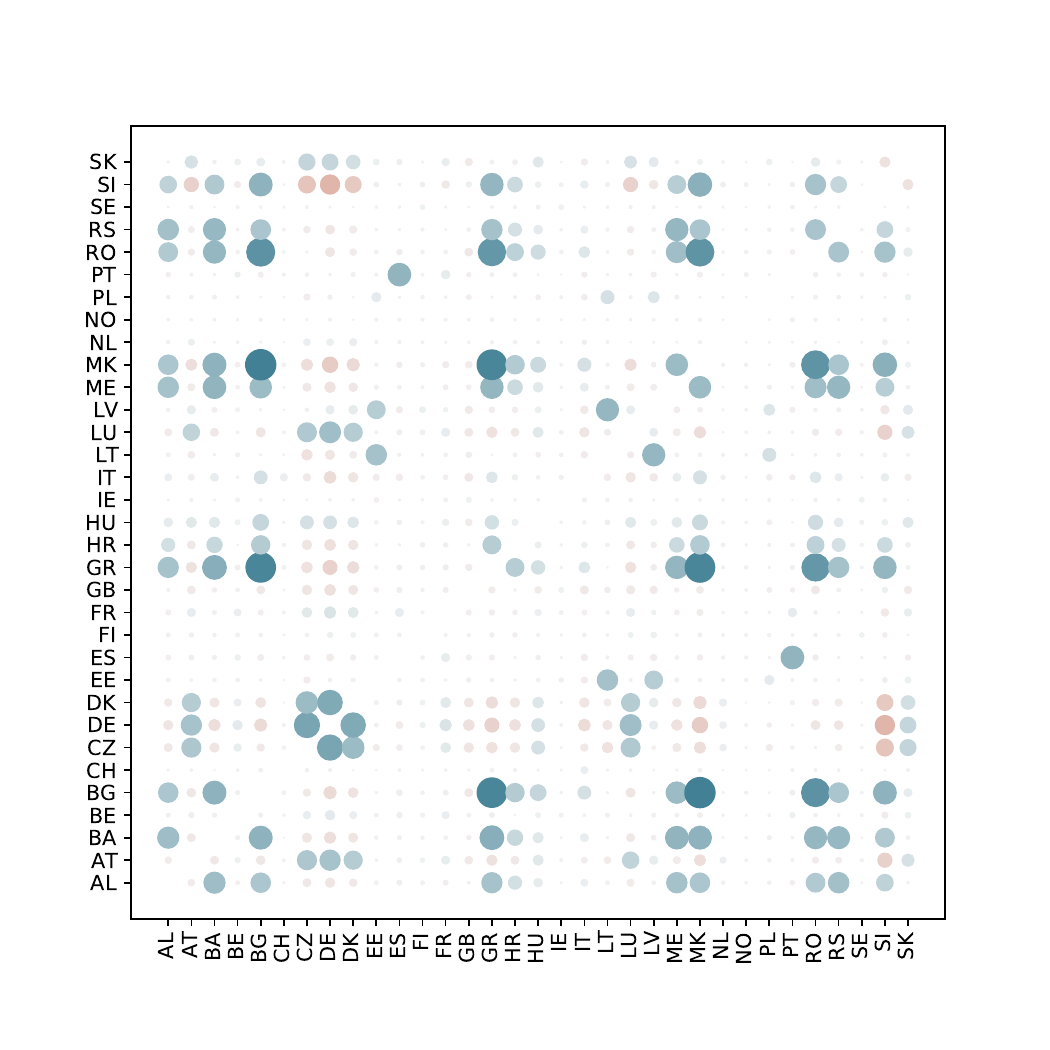}
               \caption{\textbf{Correlation of CO$_2$ abatement costs -} The data used to create this figure is the Pearson correlation of the national CO$_2$ abatement costs across all samples. a) CO$_2$ abatement cost correlations shown on a map of the model countries. Correlation strength and direction is indicated by the link color and size. Correlations below 0.2 has been removed for clarity. b) A matrix plot of CO$_2$ abatement cost correlation for all model countries. Correlation strength is indicated by marker size and color.}
               \label{fig:co2_cost_correlation}
\end{figure}

By analyzing the correlations of realized national CO$_2$ abatement cost resulting from all the configurations, Figure \ref{fig:co2_cost_correlation} is generated. From Figure \ref{fig:co2_cost_correlation} a) it is evident that for neighboring countries outcomes are closely linked. A large group of countries in the Balkans can be observed to be tightly correlated. The strong positive correlations among this group of countries indicate that the cost of abating emissions is shared among these countries. Germany is found to play a key role in the emission profiles of many central European countries. Strong positive correlations are seen between Germany and the neighbors Austria, Denmark, Luxembourg, and the Czech Republic. This indicates that in accepted scenarios where Germany is experiencing high CO$_2$ abatement cost, the neighboring countries are likely to experience higher than average costs too. A likely explanation is that energy production, and thereby also emissions, are moved from the country with tight national targets to countries with available allowable emissions. This behavior is expected when inhomogeneous carbon prices are enforced. The findings correlate well with the findings in \cite{schlott2021carbon}, where reassignment of emission reduction internally in Europe is identified as inhomogeneity of carbon prices increases. These findings are also supported by supplementary Figure \ref{fig:co2_correlation} showing correlations in CO$_2$ emissions.
A cluster of tightly correlated countries is, furthermore, found among the Baltic Sea countries of Estonia, Lithuania, and Latvia. This reveals a dynamic where CO$_2$ intensive backup energy generation moves between these countries depending on where national emission reduction targets are tightened the most. In this region of Europe renewable resources are low, and the strong correlation could indicate that the respective countries somehow are dependent on a pool of potential but dirty power exports.

\begin{figure}[htb]
               \includegraphics[width=1\textwidth,trim={1.5cm 0.6cm 2.5cm 1cm},clip]{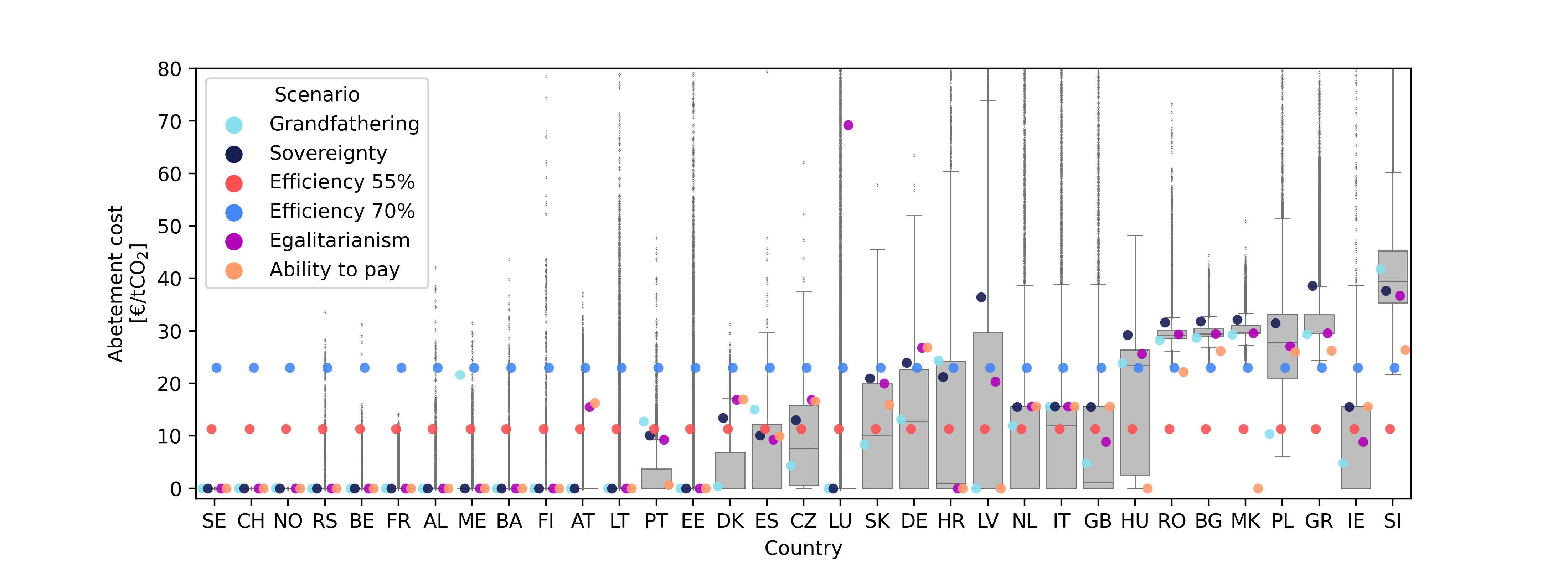}
               \includegraphics[width=1\textwidth,trim={1.5cm 0.6cm 2.5cm 1cm},clip]{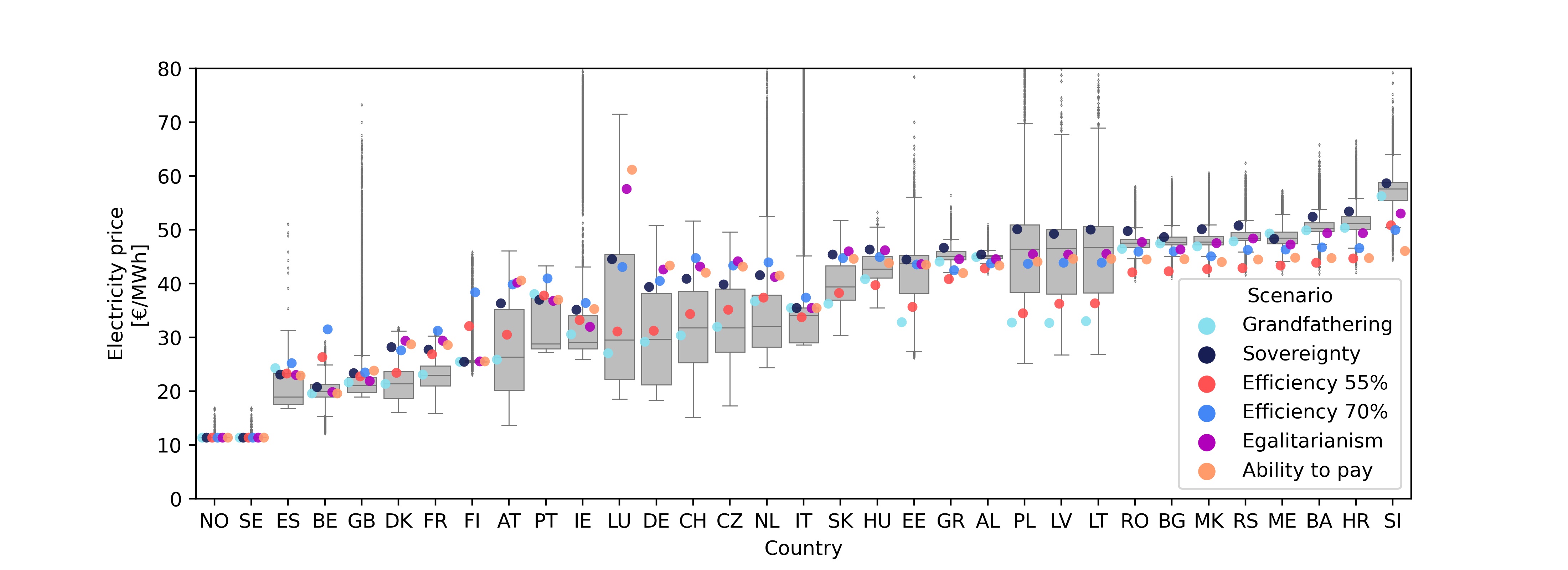}
               \caption{\textbf{National CO$_2$ abatement costs and electricity prices -} a) National CO$_2$ abatement costs for all countries across all scenarios. The countries are sorted after average emission per Mwh  produced. b) Annualy averaged hourly electricity price for the individual countries across all scenarios.}
               \label{fig:prices}
\end{figure}

Imposing a limit on national CO$_2$ emissions naturally triggers a shadow price on emissions abatement.  In a linear optimization model such as the one used in this paper, the Lagrange/Karush Kuhn-Tucker multiplier of the national CO$_2$ constraints serves as a proxy for the national CO$_2$ abatement cost. The Lagrange multiplier measures the change in system cost caused by a marginal change in the given constraint. The CO$_2$ abatement costs can be interpreted as the policy push - e.g., in the form of subsidies - required to obtain a given national emission target. High abatement cost shows that the electricity supply is hard to decarbonize. 
CO$_2$ abatement costs for all modeled configurations are shown in the top panel of Figure \ref{fig:prices}. 
Equivalently, the hourly national electricity price can be found as the Lagrange multiplier value of the national electricity balance constraints. Time-averaged electricity prices are shown on the lower panel of Figure \ref{fig:prices}. For the formulation of the linear optimization problem, including the national CO$_2$ constraint, and the nodal energy balance constraints, refer to the experimental procedures. \newline

CO$_2$ abatement costs are highly dependent on the national availability of renewable or low-carbon resources and current energy infrastructure. An abatement cost only occurs when the national emission reduction constraint is binding. Thus, in a scenario where a given country is only utilizing parts of the allocated CO$_2$ target, an abatement cost of 0 will be obtained. In Figure \ref{fig:prices}, it is seen that a large number of the model countries have CO$_2$ abatement costs of zero, indicating that these countries are not fully exploiting their assigned national emission targets. Considering the group of countries, which always utilize their allocated emissions (see Figure \ref{fig:unused_co2}), and their incurred CO$_2$ abatement costs shown in Figure \ref{fig:prices}, these countries are seen to always have non-zero abatement costs. The abatement costs for these countries are ranging from 30 to 40~€ per ton CO$_2$ in most configurations of national reduction target allocations, but with outliers ranging much higher.
On the top panel of Figure \ref{fig:prices}, the Efficiency schemes are seen having a uniform global CO$_2$ abatement cost determined by the joint emissions reduction target. The change in global CO$_2$ abatement costs between the Efficiency configurations, with 55\% and 70\% reduction, are reflecting the increased socio-economic burden associated with higher emissions reductions. 

Electricity prices are found to have a smaller spread for the individual countries as seen on the lower panel of Figure \ref{fig:prices}. The robustness of the electricity price does, however, depend on the country observed with countries at each end of the figure having more robust prices, and countries towards the center having larger deviations. The countries observed to have constantly high prices are to a large extent the same countries that had high abatement costs. The cluster of countries observed to have high variation in the electricity price is also the countries observed to have strong correlations in national CO$_2$ emissions in Figure \ref{fig:co2_correlation}. The observed electricity prices on the lower panel of Figure \ref{fig:prices} span from 10 €/MWh to above 70 €/MWh for some outlier outcomes. This span in power prices is rather large compared to current power prices which are around 50 €/MWh for most European countries.

\begin{figure}[htb]
               \includegraphics[width=1\textwidth,trim={0 0 0 0 },clip]{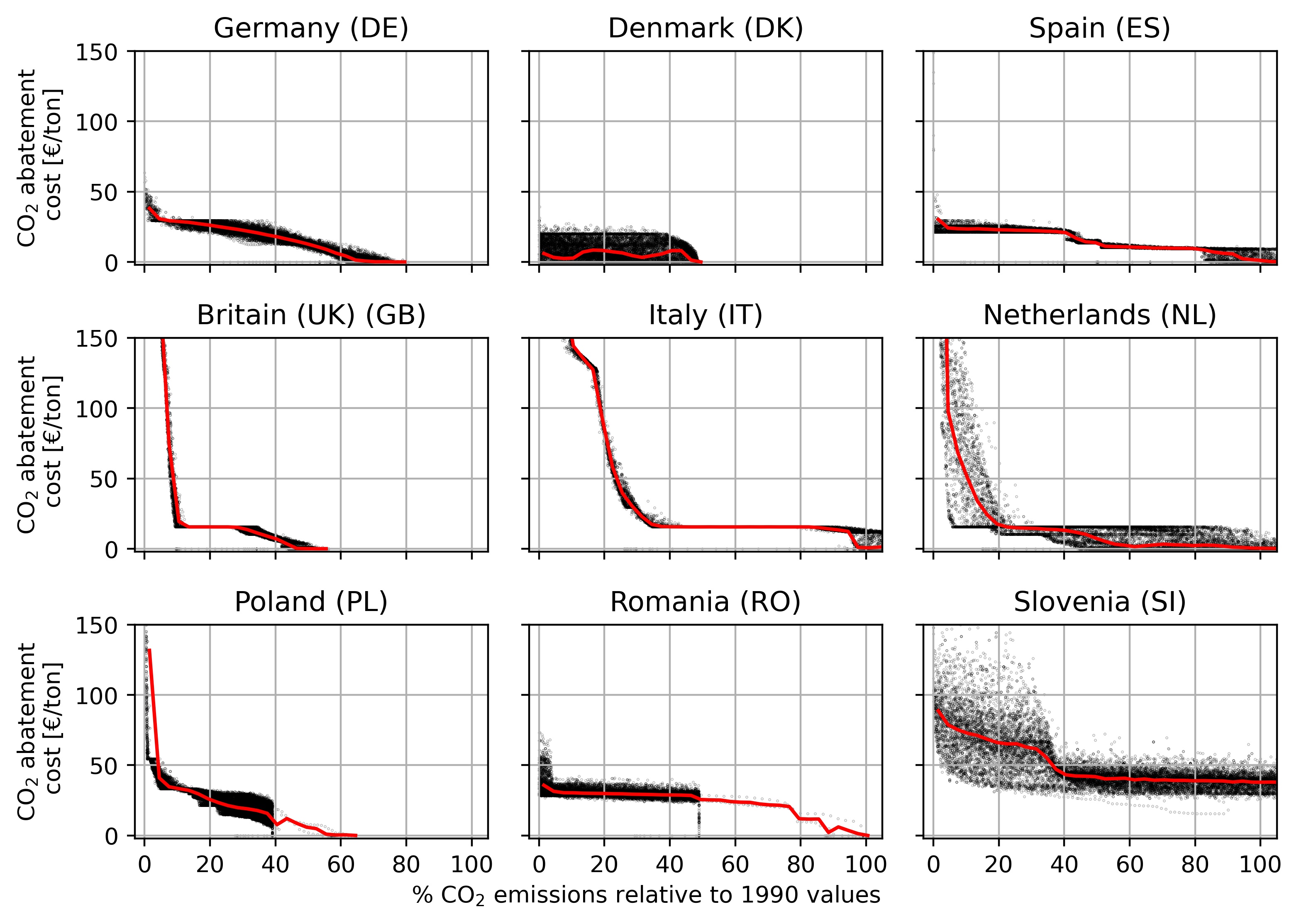}
               \caption{\textbf{CO$_2$ abatement cost dependence on emission level - } The figure shows CO$_2$ abatement costs for a select group of model countries. The individual samples are shown with black dots and the mean with a red line. The figure is available for all model countries at \ref{fig:all_abatement_price}. }
               \label{fig:abetement}
\end{figure}

The price of abating CO$_2$ emissions is naturally depending on the emission reduction level. Figure \ref{fig:abetement} show the reliance of CO$_2$ abatement cost on the emission reduction level for a select number of countries. CO$_2$ abatement cost is observed to be tightly correlated with the emission reduction level in a highly nonlinear manner. In the figure, some countries can be observed to have abatement costs increasing dramatically as their reduction levels approach zero. Other countries are found to have a constantly gradually increasing abatement cost as emissions are being reduced. Analyzing the composition of the energy supply of these countries in supplementary Figures \ref{fig:all_renewable_cap}, \ref{fig:all_storage_cap}, and \ref{fig:all_OCGT_cap}, the countries where abatement costs increase drastically at low emissions can be observed to install large shares of renewables at these reduction levels. Thus the increase in abatement cost is driven by a drastically increasing need for renewable energy sources.
In the special case of Denmark (DK), the mean abatement cost can be observed to fluctuate with decreasing emissions. The fluctuations in mean abatement cost seen are caused by the finite number of samples used in this work. There is, furthermore, a noticeable difference between when the countries obtain an abatement cost of 0 €/ton, ranging from 50\% to 100\% CO$_2$ emissions relative to 1990 values. CO$_2$ abatement of 0 means that it is cost-optimal to reduce emissions beyond this point. In Great Britain (GB) it would, e.g., be cost-optimal to reduce emissions below 60\% of 1990 levels.

Studying how the five CO$_2$ allocation approaches from Table \ref{tab:scenarios} are distributed in supplementary Figure \ref{fig:unused_co2}, reveals that the Efficiency approach ensures that all national emission quotas are utilized 100\%. In the Efficiency scheme, no country is assigned more emissions than needed, whereas the other allocation approaches result in inefficient allocations increasing total system cost. Especially the Grandfathering scheme leads to a large share of assigned emissions being left unused. It should be noted, that as only the electricity sector is considered in this research, the 55\% reduction target provides a somewhat conservative benchmark, considering that the electricity sector is expected to be the first to be decarbonized. As several countries already have decarbonized their electricity sector to a high extent, they can be expected to find it economically optimal to avoid using all the emissions they are assigned. Norway, Sweden, and Switzerland can be seen to refrain from using their allocated emissions in most configurations of national reduction target allocations. This is in accordance with the findings shown in Figure \ref{fig:co2_mwh_box}.

\section{Discusion}
What has been considered here is possible effort-sharing configurations for the European power sector to reach a joint decarbonization target of 55\%, relative to 1990. This is not to be confused with strict quota allocation schemes, as the perspective here is broader. As such, the scenarios investigated here do not represent alternatives to the EU ETS, but should rather be seen as an investigation of the financial motivation required in each individual country to obtain certain effort-sharing configurations.

The national CO$_2$ abatement costs associated with varying levels of national decarbonization have been identified. The CO$_2$ abatement cost can be interpreted as the policy push required to obtain a given national emission reduction target. This policy push could, in practice, be implemented in a number of ways. Effective examples for incentivizing emission reductions include subsidies for renewable technologies, publicly funded projects, and performance standards \cite{policy101}. In this work, the magnitude of the policy push is determined but the specific policy tool is not discussed. 

The results of this work show how European countries react very differently to changing national emission targets, revealing large inequality in the efforts required to decarbonize national electricity supplies. Where some nations find it cost-optimal to reduce emissions beyond the 55\% ambition, other nations require a strong policy push to achieve the same goal. Under a global CO$_2$ price - such as EU ETS - it will be cost-optimal to continue reliance on fossil fuel-based electricity production in countries with hard to decarbonize electricity supplies. If the inequality in decarbonization costs is not addressed, these countries will be left behind in the transformation of the European energy sector. Some of the countries where decarbonization of the electricity sector has been found especially costly include Slovenia, Greece, Poland, North Macedonia, Bulgaria, and Romania. Following the sovereignty effort sharing principle would result in a more equitable distribution of CO$_2$ intensity, but it will drive higher CO$_2$ abatement costs in Poland and other countries with high emission intensities. 
An overlap can be observed between the group of countries experiencing high abatement costs and the countries experiencing high electricity prices. Countries included in both groups include several Eastern European and Balkan countries. As these countries also have a tendency to have high-income inequality, increasing electricity prices may significantly grow energy poverty. Thus, decarbonization governed solely by a global CO$_2$ price may drastically increase social and economic inequality.

EU ETS is not equitable on its own. EU ETS is currently governing emission prices in all EU countries \cite{eu_ets}. The price of emission permits is the same across Europe, despite differences in GDP, income, and purchasing power. Countries with low incomes tend to have inefficient and dirty power sectors, further reinforcing the inequities, as the costs to shoulder decarbonization tend to be relatively larger for them. To ensure equity under EU ETS financial support must be provided to nations where CO$_2$ abatement costs are highest. 
The commission proposes to create an ETS2 for transport and households, hence it is relevant to explore effort-sharing principles and their implications. Especially so as the Just Transition Fund \cite{just_transition_mechanism} is relatively loosely described.

A fully economically efficient outcome is hard to obtain, as it requires close coordination and full information. Although in theory, EU ETS should ensure an efficient outcome, in practice information asymmetries are distorting the clearing of markets. Moreover, there are sunk costs causing dirty power producers to continue operating, due to the differences between marginal and average costs of production.
The results obtained in this work show how small deviations from the cost-optimal effort sharing configuration will lead to increases in total system cost. 

The differences in abatement costs per unit of carbon can be misleading as to the distributional consequences, as the abatement effort will be less demanding in countries that are already partly decarbonized such as France. Still, not only can unit costs be expected to be relatively high in Balkan countries, but these countries are also facing a greater transformation overall, involving high absolute costs. Thus, it is important to consider the volume of emissions that must be abated along with the marginal abatement cost. 

The allocation principle of sovereignty results in the highest cost overshoot, showing the value of European cooperation. Still, the fluidity of the transactions in the power market and the facilitation of more interconnectors are making planning for decarbonization challenge. Some countries in the center of Europe are facing many different feasible outcomes, whereas countries in the periphery (Finland) will have fewer choices to make. The vision of a fully interconnected Europe is still only a distant dream.

\section{Conclusion}

A global price on CO$_2$ emissions ensures economically efficient decarbonization of the European electricity sector but fails to provide a just energy transition. This paper investigates the spread in national costs associated with decarbonizing the electricity supply in a European context. By performing an MCMC analysis of national emission reduction targets using the adoptive Metropolis-Hastings method, in combination with an energy system optimization model, the inequality in the cost of decarbonization is revealed.

The national costs of abating emissions identified in this study reveal large inequalities among the European countries. While some nations find it cost-optimal to reduce emissions to fulfill their part of global commitments, other nations must implement large financial incentives for energy producer to reduce their emissions. The cost associated with supporting the green transition with financial support naturally entails a loss in national welfare as resources that could have otherwise been prioritized for other parts of society has to be used to reduce emissions from the energy sector. The highest abatement costs are found in nations with relatively low GDPs. Thus, if all European nations are expected to contribute equally, in terms of emission reductions, there is a potential for severely increased inequality among EU nations. Only through the strong financial support of the nations with hard to decarbonize electricity sectors can a just transition be achieved. 
High CO$_2$ abatement costs have been observed to correlate with high electricity prices. This further increases the risk of inequality in the transition. A range of low GDP countries has been observed to have high abatement costs and thereby also potentially high electricity prices. Increasing electricity prices in a low GDP country has a great risk of increasing energy poverty.
The range of possible national CO$_2$ reduction targets identified for the European electricity supply reveals that a cost increase of 5\% from the cost-optimal solution is almost inevitable. Small deviations from the cost-optimal configuration of national CO$_2$ reduction targets will lead to increases in total system cost. 
A strong correlation between CO$_2$ abatement costs within three groups of countries located in central, North Eastern, and South-Eastern Europe was identified. These correlations reveal that enforcing a tight CO$_2$ limit on one cluster country will move emissions, and thereby also abatement costs, to the other countries in the cluster. The likely explanation for this behavior is that carbon leakage between neighboring countries will occur when differences in national CO$_2$ abatement cost increase. 
To ensure equity in the transition towards carbon neutrality of the European energy sector the nations with easy to decarbonize energy sectors must provide financial support to nations where emission reductions are costly. In a system with a homogeneous price on CO$_2$ emissions, additional policy is required to distribute resources if equal effort sharing is to be achieved. 

\section{Experimental procedures}
\subsection{Resource availability}

\textbf{Lead contact}\\
Tim Pedersen, Email: ttp@mpe.au.dk\\

\textbf{Description}\\
The developed python scripts used in the paper are available under an open-source license. All scripts and data required to reproduce the results from this paper are openly available.\\

\textbf{License}\\
The software is under the GNUv3 license\\

\textbf{Data and code availability}\\
Link to the repository will be public upon completion of review\\

\subsection{Energy system optimization model}\label{apen:model}
The joint capacity and dispatch energy system optimization model used in this work is based on the PyPSA-Eur-Sec model \cite{PyPSA-EUR-SEC}. The PyPSA-Eur-Sec model to a high extent depends on data imports from the PyPSA-Eur model \cite{horsch2018pypsa}. The model formulated in this work represents a 2030 brownfield scenario of the European electricity supply spanning 33 ENTSO-E member countries, i.e. the model includes EU-27 without Cyprus and Malta, instead including Norway, Switzerland, Serbia, Bosnia-Herzegovina, Albania, Montenegro, Macedonia and, United Kingdom. 

A brownfield scenario is generated where existing capacities that are planned to be in operation by 2030 are included in the model. The included brownfield capacities are seen in supplementary Table \ref{tab:brownfield_cap}. Existing conventional capacities are found from the power plant matching database \cite{powerplantmatching}, while renewable capacities are found from the IRENA annual statistics \cite{IRENA}. A minimum requirement of 55\% CO$_2$ reductions has been used throughout this work, corresponding to an annual CO$_2$ budget of 666.85 Mton CO$_2$. All existing plus the planned transmission capacities in the Ten Year Network Development Plan (TYNDP) \cite{tyndp_2018} are included. Transmission capacities are seen in Figure \ref{fig:starting_point}.

Technology capacities can be expanded to meet energy demand. Cost of the expandable technologies are given in supplementary Table \ref{tab:tech_cost}. Efficiency and emission data are available in supplementary Table \ref{tab:tech_data}. Technology costs are primarily based on the 2030 cost prediction given by the Danish Energy Agency in their technology data catalog \cite{Data_catalogue}. A discount rate of 7\% has been used to calculate annualized costs using the annuity factor given in Equation \ref{eq:annuity}. Here $r$ is the discount rate and $n$ is the technology lifetime. 

\begin{equation}\label{eq:annuity}
    a = \frac{1 - (1+r)^{-n}}{r}
\end{equation}

The model of the European power sector is formulated as a linear optimization problem, consisting of an objective function along with a set of constraints. Throughout this description of the model, the model variables are split in two vectors namely $\mathbf{x}$ and $\mathbf{y}$. Where $\mathbf{x}$ describes the national CO$_2$ reduction target given by the MCMC sampler $\mathbf{x} = {r_n \; \forall \; n}$. Here $r_n$ is the national CO$_2$ target in tons CO$_2$ for all model countries $n$. The remaining variables $\mathbf{y}$ represent technology capacities and dispatch $\mathbf{y} = \{ \mathbf{g}_{n,s,t} , \mathbf{G}_{n,s} , \mathbf{F}_{l} \}$. 
Here index $s$ is indexing the technology for all technologies included in the model, index $t$ is indexing the hour for all hours in the year, and $l$ represents the transmission line. The variables determined in the optimization process are thus:  

\begin{itemize}
	\item $\mathbf{g}_{n,s,t}$ : Hourly dispatch of energy from the given plants in the given countries with the marginal cost $\mathbf{o}_{n,s}$.
	\item $\mathbf{G}_{n,s}$: Total installed capacity of the given technologies in the given countries with the capital cost $\mathbf{c}_{n,s}$.
	\item $\mathbf{F}_{l}$: Total installed transmission capacity for all lines with the fixed annualized capacity cost $\mathbf{c}_{l}$.
\end{itemize}

The model is then formulated as a linear problem following the standard formulation given as:

\begin{equation}\label{eq:ConvexOptimization}
\begin{split}
\text{minimize} \;&\; \mathbf{f}_0(\mathbf{y})   \; \\
\text{subject to} \; &\; \mathbf{f}_i(\mathbf{x,y}) \leq 0 \; \; i=1..m\\
\;            &\;  \mathbf{h}_i(\mathbf{x,y}) = 0 \; \; i=1..p\\
\end{split}
\end{equation}

The national CO$_2$ targets $\mathbf{x}$ are given by the MCMC sampler and are thus not optimized in the model. Only the technical variables $\mathbf{y}$ are optimized in the optimization problem. 

The objective function of the model is to minimize total system cost and can be formulated as follows: 

\begin{equation}
\text{minimize} \; f_0(\mathbf{x,y}) =  \sum_{n,s} \mathbf{c}_{n,s} \mathbf{G}_{n,s} + \sum_l \mathbf{c}_l \mathbf{F}_l + \sum_{n,s,t} \mathbf{o}_{n,s} \mathbf{g}_{n,s,t} 
\end{equation}

The model assumes perfect competition and foresight as well as long-term market equilibrium. 
For all model nodes and all hours in the year, a power balance constraint is enforced requiring that the energy demand $\mathbf{d}_{n,t}$ is fulfilled.
Energy demand data is taken from the ENTSO-E data portal \cite{ENTSOE} and decomposed into industrial and residential demand following the method given in \cite{horsch2018pypsa}. 
The incidence matrix describing the line connections is given by $\mathbf{K}_{n,l}$ and the hourly power flowing through each line is described as $\mathbf{f}_{l,t}$. The nodal power balance constraint can then be formulated as:

\begin{equation} \label{eq:equality_constraint}
\sum_s \mathbf{g}_{n,s,t} - \mathbf{d}_{n,t} - \sum_l \mathbf{K}_{n,l} \mathbf{f}_{l,t} = 0  \; \; \forall n,t
\end{equation}

The dispatch of each technology $\mathbf{g}_{n,s,t}$ is limited by the installed technology capacity $\mathbf{G}_{n,s}$. The dispatch of renewable energy generators such as wind and solar are furthermore limited by the hourly capacity factor $\mathbf{\overline{g}}_{n,s,t}$. The capacity factor for conventional power plants is 1, whereas it is generated from weather data for the renewable generators. A detailed explanation of the derivation of renewable generation potentials is given in \cite{horsch2018pypsa}.

\begin{equation}
0 \leq \mathbf{g}_{n,s,t} \leq \mathbf{\overline{g}}_{n,s,t} \mathbf{G}_{n,s} \; \forall n,s,t
\end{equation}

Similarly, the power $\mathbf{f}_{l,t}$ flowing through the transmission lines is also limited by the installed capacity. As the direction of the transmission is without significance it is the absolute transmission $|\mathbf{f}_{l,t}| $ that is limited.  

\begin{equation}
|\mathbf{f}_{l,t}| \leq \mathbf{F}_l \; \forall l,t
\end{equation}

The maximum capacity allowed for each technology is determined by geographical potentials available $\mathbf{G}_{n,s}^{max}$. 

\begin{equation}
0 \leq \mathbf{G}_{n,s} \leq \mathbf{G}_{n,s}^{max} \; \forall n,s
\end{equation}

CO$_2$ emissions can be constrained in two ways. Either through a global constraint on emissions, or by national constraints on emissions. The global CO$_2$ reduction constraint is formulated as:

\begin{equation}\label{eq:global_co2}
\sum_{n,s,t} \frac{1}{\mathbf{\eta}_s}\mathbf{g}_{n,s,t} \mathbf{e}_s -CAP_{CO_2} \leq 0 \; 
\end{equation}

Here the $CAP_{CO_2}$ is the global emissions limit given in ton CO$_2$. $\mathbf{\eta}_s$ is the generator efficiency and $\mathbf{e}_s$ is the CO$_2$ equivalent emission intensity of the fuel. Note that only a single constraint is given here. Limiting emissions through national constraints can be done by defining a constraint for each country in the model. The national emissions targets $\mathbf{x}$ are given by the MCMC sampler. 

\begin{equation}\label{eq:national_co2}
\sum_{s,t} \frac{1}{\mathbf{\eta}_s}\mathbf{g}_{n,s,t} \mathbf{e}_s - \mathbf{r}_i \leq 0 \; 
\end{equation}

The global CO$_2$ constraint (Equation \ref{eq:global_co2}) is only used in the Efficiency scenario. In all other scenarios, the national CO$_2$ targets are explicitly given, either by the sampler or following a certain allocation scheme. 

When the model is solved the Lagrange multipliers associated with every constraint are also obtained as an output. The value of these Lagrange multipliers represents the cost increase/decrease associated with tightening/loosening the constraint by one unit. Thus by evaluating the Lagrange multipliers associated with the energy balance constraint (Equation \ref{eq:equality_constraint}) the nodal hourly electricity price can be obtained. Similarly, the Lagrange multiplier of the national CO$_2$ target constraint (Equation \ref{eq:national_co2}) provides a proxy for the national CO$_2$ abatement cost. \\

\subsection{Sampling method}\label{apen:sampler}
To draw possible CO$_2$ target configurations the Adaptive Metropolis-Hastings (AMH) sampler is implemented \cite{haario_2001}. The AMH sampler is based on a Markov Chain process where samples are continuously drawn from a proposal distribution centered around the previous sample point. By controlling the width of the proposal distribution continuously, the AMH sampler ensures efficient sampling. The AMH sampler is chosen as it is simple to implement while providing efficient sampling and fast mixing \cite{solonen2012efficient}. 


An arbitrary CO$_2$ target configuration can be denoted as the vector $\mathbf{x}$, with each component of this vector $x_i$ representing the national CO$_2$ emission target of the $i$'th country relative to the total CO$_2$ emission target. The allowed emission for a given country can be determined as $x_i \cdot CO2_{CAP}$, where the $CO2_{CAP}$ is the total global amount of CO$_2$ emissions allowed in tonnes of CO$_2$. Realizations of the variables are denoted with subscript $\mathbf{x}_t$. It is important to note that the sum of $\mathbf{x}$ can be greater than 1, and thus the combined emission targets can add up to more than the total global amount of CO$_2$ emissions allowed $CO2_{CAP}$. As the rejection criteria are based on the realized emissions, a sample where the sum of $\mathbf{x}$ is greater than 1 can be accepted if not all allowed emissions are realized. Similarly $\mathbf{x}$ can also sum to less than 1. In such a case the global emission reductions will be less than the required 55\% and the sample can be immediately rejected.  

Given a starting point $\mathbf{x}_0$ the AMH sampler will continuously generate new sample proposals $\mathbf{x}'$. New samples are drawn from the proposal distribution centered around the previous sample. The proposal distribution is defined as a uniform distribution around the previous sample point with the width $\sigma$. Thus the maximal change in each variable $x_i$ per iteration is $\sigma/2$. There are however a few caveats. As the variables considered $\mathbf{x}$ are fractions of a total CO$_2$ budget, they are constrained to be between 0 and 1. Therefore, the uniform distribution is bounded not to exceed this area.
The starting point $\mathbf{x}_0$ used is the cost-optimal scenario also denoted as the Efficiency scenario. A burn-in period of 100 samples is used by discarding the first 100 samples of each chain to remove any bias towards the starting point. 

\begin{equation}
    \mathbf{x}' \sim \mathcal{U}[\max(\mathbf{x}_{t-1}-\frac{\sigma}{2} ,0) , \min(\mathbf{x}_{t-1}+\frac{\sigma}{2} ,1) ]
\end{equation}

The distribution width $\sigma$ is tuned continuously as more information about the solutions space is obtained. By setting $\sigma$ too low, the sampler will need an excessive amount of samples to explore the entire solution space. On the other hand, setting $\sigma$ too high will result in the rejection of too many samples. 
By continuously monitoring the acceptance rate, it is possible to determine if the chain is taking either too short or long steps. If the acceptance rate is very high $\sigma$ should be increased, and if the acceptance rate is low $\sigma$ should be decreased. In practice, this is implemented by letting the sampler run for a number of iterations and evaluating the acceptance rate in that batch of samples. In this implementation of the AMH sampler, $\sigma$ is updated by continuously monitoring the acceptance ratio of the samples. When the acceptance ratio is below a user-specified value, $\sigma$ is incremented by a small amount $\epsilon$, and vice versa when the acceptance ratio is too high. An $\epsilon$ value of 0.05  and a desired acceptance ratio of 80\% have been used throughout this work. 

The feasibility of a proposed sample $\mathbf{x}'$ is evaluated using the energy system optimization model. If the solution to the energy system optimization model given $\mathbf{x}'$ as input satisfies all criteria from Table \ref{tab:feasa}, the sample is accepted. Otherwise, the sample is rejected and a new proposal sample is drawn. When a proposed sample is accepted it is assigned index $t$, such that $\mathbf{x}_t = \mathbf{x}'$. If a sample is rejected the previous sample point is stored instead $\mathbf{x}_t = \mathbf{x}_{t-1}$. 
The process of drawing samples from the proposal distribution and either accepting or rejecting them is repeated until a sufficient sample size is reached. 

The result is a set of realizations of $\mathbf{x}$ that can ensure feasible operation of the model, global emission reductions higher than the base scenario, and a total system cost that is no more than 18\% higher than that of the base scenario. If enough samples are drawn the distribution of the set of realizations will approximate all solutions satisfying the above-mentioned criteria. 

In practice, the above algorithm is implemented as a parallel process with multiple chains running simultaneously. The samples from the parallel chains can then be merged at the end of the sampling process. 

\subsection{Principle effort sharing schemes}
Five principles for allocation of national reduction targets have been used as reference. Inspired by Zhou et al. \cite{Zhou_2016}, these are grandfathering, sovereignty, efficiency, egalitarianism, and ability to pay. The procedures for the allocation of national reduction targets, and conversely the emissions, for each of these five principles, are shown in Table \ref{tab:scenarios}. Using these five principles, six emission reduction configurations have been created as seen in Supplementary Figure \ref{fig:co2_scenarios}. The Efficiency 55\% and Efficiency 70\%, configurations correspond to using EU ETS at 55 and 70\% joint reductions respectively, whereas, the Grandfathering, Sovereignty, Egalitarianism and, Ability to pay configurations represent alternatives to EU ETS. 
These configuration principles are implemented such that they all distribute the same CO$_2$ budget, except for the efficiency 70\% reduction configuration. As allowable emissions are left unused by some countries, as they find it economically favorable to do so, the total realized CO$_2$ reduction for all configuration principles other than Efficiency, will be higher than the minimum goal of 55\%. The configuration principles could alternatively be implemented to all have realized emissions corresponding to a 55\% CO$_2$ reduction. A choice was made to use configurations with equal CO$_2$ budgets rather than equal realized emissions, as they represent a more diverse set of scenarios and better represent a decision process where budgets are allocated as national targets that will be realized, e.g. a decade later. For results using configurations with equal realized emissions see \ref{sec:all55scenarios}.

\subsection{Methodology limitations}
Weather and demand patterns are expected to change as a result of global warming and general electrification of energy use. Investigating these effects is, however, beyond the scope of this paper.
Only the electricity sector has been modeled. At a European level, the effects of sector coupling are only expected to be moderate by 2030, thus, this simplification is believed to provide only a minor source of error. If sector coupling was implemented, the electricity sector could be expected to achieve a higher decarbonization rate than involved with the 55\% target, as it is considered easier to achieve here than in other sectors. 

\section{Acknowledgments}
Financial support (grant \#82841) from Nordforsk, Nordic Energy and Nordic Innovation to the project NOWAGG (New Nordic Ways to Green Growth;). Tim T. Pedersen, M. Victoria nad Gorm B. Andresen are fully or partially funded by the RE-INVEST project, which is supported by the Innovation Fund Denmark under grant number 6154-00022B. Tim T. Pedersen and Gorm B. Andresen are partialy fundet by DFF-Dansk ERC-støtteprogram (case number: 1105-00001B). \newline

Assistance and input given by Martin Greiner was greatly appreciated. 

\section{Author contributions}
Conceptualization T.T.P, M.V. and G.B.A; Methodology T.T.P; Software T.T.P; Resources M.V.P and G.B.A; Writing - Original Draft T.T.P. and M.S.A; Writing - Review \& Editing T.T.P, M.S.A, M.V, and G.B.A; Funding Acquisition G.B.A;

\section{Declaration of competing interest}
The authors declare no competing interests.

\clearpage
\bibliography{main.bib}

\begin{thebibliography}{10}
\expandafter\ifx\csname url\endcsname\relax
  \def\url#1{\texttt{#1}}\fi
\expandafter\ifx\csname urlprefix\endcsname\relax\def\urlprefix{URL }\fi
\expandafter\ifx\csname href\endcsname\relax
  \def\href#1#2{#2} \def\path#1{#1}\fi

\bibitem{nordhaus2019}
W.~Nordhaus, Climate change: The ultimate challenge for economics, American
  Economic Review 109~(6) (2019) 1991--2014.
\newblock \href {https://doi.org/DOI: 10.1257/aer.109.6.1991} {\path{doi:DOI:
  10.1257/aer.109.6.1991}}.

\bibitem{green_deal}
Secretariat-General,
  \href{https://eur-lex.europa.eu/legal-content/EN/ALL/?uri=COM:2019:640:FIN}{Communication
  from the commission - the european green deal} (2019).
\newline\urlprefix\url{https://eur-lex.europa.eu/legal-content/EN/ALL/?uri=COM:2019:640:FIN}

\bibitem{van2020implications}
N.~J. Van~den Berg, H.~L. van Soest, A.~F. Hof, M.~G. den Elzen, D.~P. van
  Vuuren, W.~Chen, L.~Drouet, J.~Emmerling, S.~Fujimori, N.~H{\"o}hne, et~al.,
  Implications of various effort-sharing approaches for national carbon budgets
  and emission pathways, Climatic Change 162~(4) (2020) 1805--1822.

\bibitem{NDC_report}
U.~Secretariat, \href{https://unfccc.int/documents/306848}{Nationally
  determined contributions under the paris agreement. synthesis report by the
  secretariat} (2021).
\newline\urlprefix\url{https://unfccc.int/documents/306848}

\bibitem{tol2009economic}
R.~S. Tol, The economic effects of climate change, Journal of economic
  perspectives 23~(2) (2009) 29--51.

\bibitem{hardin1968tragedy}
G.~Hardin, The tragedy of the commons: the population problem has no technical
  solution; it requires a fundamental extension in morality., science
  162~(3859) (1968) 1243--1248.

\bibitem{Bauer_2020}
N.~Bauer, C.~Bertram, A.~Schultes, D.~Klein, G.~Luderer, E.~Kriegler, A.~Popp,
  O.~Edenhofer, Quantification of an efficiency–sovereignty trade-off in
  climate policy, Nature 588~(7837) (2020) 261–266.
\newblock \href {https://doi.org/10.1038/s41586-020-2982-5}
  {\path{doi:10.1038/s41586-020-2982-5}}.

\bibitem{just_transition_mechanism}
The just transition mechanism (2021).

\bibitem{jenkins_2016}
K.~Jenkins, D.~McCauley, R.~Heffron, H.~Stephan, R.~Rehner, Energy justice: a
  conceptual review, Energy Research \& Social Science 11 (2016) 174--182.
\newblock \href {https://doi.org/https://doi.org/10.1016/j.erss.2015.10.004}
  {\path{doi:https://doi.org/10.1016/j.erss.2015.10.004}}.

\bibitem{maguire_2021}
D.~Maguire, C.~Shaw, Fair energy transition for all - literature review (2021).

\bibitem{victoria_2020}
M.~Victoria, K.~Zhu, T.~Brown, G.~B. Andresen, M.~Greiner, Early
  decarbonisation of the european energy system pays off, Nature communications
  11~(1) (2020) 1--9.
\newblock \href {https://doi.org/https://doi.org/10.1038/s41467-020-20015-4}
  {\path{doi:https://doi.org/10.1038/s41467-020-20015-4}}.

\bibitem{eurostat_2019}
Eurostat, Greenhouse gas emissions by source sector (2019).

\bibitem{solomon2011coming}
B.~D. Solomon, K.~Krishna, The coming sustainable energy transition: History,
  strategies, and outlook, Energy Policy 39~(11) (2011) 7422--7431.

\bibitem{dyrhauge2017denmark}
H.~Dyrhauge, Denmark: a wind-powered forerunner, in: A guide to EU renewable
  energy policy, Edward Elgar Publishing, 2017.

\bibitem{levenda_2021}
A.~Levenda, I.~Behrsin, F.~Disano, Renewable energy for whom? a global
  systematic review of the environmental justice implications of renewable
  energy technologies, Energy Research \& Social Science 71 (2021) 101837.
\newblock \href {https://doi.org/https://doi.org/10.1016/j.erss.2020.101837}
  {\path{doi:https://doi.org/10.1016/j.erss.2020.101837}}.

\bibitem{golubchikov_2020}
O.~Golubchikov, K.~O'Sullivan, Energy periphery: Uneven development and the
  precarious geographies of low-carbon transition, Energy and Buildings 211
  (2020) 109818.
\newblock \href {https://doi.org/https://doi.org/10.1016/j.enbuild.2020.109818}
  {\path{doi:https://doi.org/10.1016/j.enbuild.2020.109818}}.

\bibitem{bohringer2009eu}
C.~B{\"o}hringer, T.~F. Rutherford, R.~S. Tol, The eu 20/20/2020 targets: An
  overview of the emf22 assessment, Energy economics 31 (2009) S268--S273.

\bibitem{grunewald2017renewable}
P.~Grunewald, Renewable deployment: Model for a fairer distribution, Nature
  Energy 2~(9) (2017) 1--2.

\bibitem{sovacool2015energy}
B.~K. Sovacool, M.~H. Dworkin, Energy justice: Conceptual insights and
  practical applications, Applied Energy 142 (2015) 435--444.

\bibitem{Zhou_2016}
P.~Zhou, M.~Wang, Carbon dioxide emissions allocation: A review, Ecological
  Economics 125 (2016) 47–59.
\newblock \href {https://doi.org/10.1016/j.ecolecon.2016.03.001}
  {\path{doi:10.1016/j.ecolecon.2016.03.001}}.

\bibitem{alcaraz_2018}
O.~Alcaraz, P.~Buenestado, B.~Escribano, B.~Sureda, A.~Turon, J.~Xercavins,
  Distributing the global carbon budget with climate justice criteria, Climatic
  change 149~(2) (2018) 131--145.
\newblock \href {https://doi.org/https://doi.org/10.1007/s10584-018-2224-0}
  {\path{doi:https://doi.org/10.1007/s10584-018-2224-0}}.

\bibitem{markowitz2012climate}
E.~M. Markowitz, A.~F. Shariff, Climate change and moral judgement, Nature
  Climate Change 2~(4) (2012) 243--247.
\newblock \href {https://doi.org/https://doi.org/10.1038/nclimate1378}
  {\path{doi:https://doi.org/10.1038/nclimate1378}}.

\bibitem{Schwenk_2020}
L.~J. Schwenk-Nebbe, M.~Victoria, G.~B. Andresen, M.~Greiner, Co2 quota
  attribution effects on the european electricity system comprised of
  self-centred actors, SSRN Electronic Journal (2020).
\newblock \href {https://doi.org/https://doi.org/10.2139/ssrn.3689207}
  {\path{doi:https://doi.org/10.2139/ssrn.3689207}}.

\bibitem{sasse2019distributional}
J.-P. Sasse, E.~Trutnevyte, Distributional trade-offs between regionally
  equitable and cost-efficient allocation of renewable electricity generation,
  Applied Energy 254 (2019) 113724.

\bibitem{drechsler2017efficient}
M.~Drechsler, J.~Egerer, M.~Lange, F.~Masurowski, J.~Meyerhoff, M.~Oehlmann,
  Efficient and equitable spatial allocation of renewable power plants at the
  country scale, Nature Energy 2~(9) (2017) 1--9.

\bibitem{haario_2001}
H.~Haario, E.~Saksman, J.~Tamminen, et~al., An adaptive metropolis algorithm,
  Bernoulli 7~(2) (2001) 223--242.
\newblock \href {https://doi.org/https://doi.org/10.2307/3318737}
  {\path{doi:https://doi.org/10.2307/3318737}}.

\bibitem{PyPSA-EUR-SEC}
T.~Brown, martavp, lisazeyen, M.~Maria, Leon, F.~Neumann, Pypsa/pypsa-eur-sec:
  Pypsa-eur-sec version 0.4.0 (2020).
\newblock \href {https://doi.org/10.5281/zenodo.4317529}
  {\path{doi:10.5281/zenodo.4317529}}.

\bibitem{EU_2030_Climate_target_plan}
2030 climate target plan (2021).

\bibitem{brill1982}
E.~D. Brill~Jr, S.-Y. Chang, L.~D. Hopkins, Modeling to generate alternatives:
  The hsj approach and an illustration using a problem in land use planning,
  Management Science 28~(3) (1982) 221--235.
\newblock \href {https://doi.org/https://doi.org/10.1287/mnsc.28.3.221}
  {\path{doi:https://doi.org/10.1287/mnsc.28.3.221}}.

\bibitem{neumann2021broad}
F.~Neumann, T.~Brown, Broad ranges of investment configurations for renewable
  power systems, robust to cost uncertainty and near-optimality, arXiv (2021).
\newblock \href {https://doi.org/https://arxiv.org/abs/2111.14443}
  {\path{doi:https://arxiv.org/abs/2111.14443}}.

\bibitem{lombardi2020policy}
F.~Lombardi, B.~Pickering, E.~Colombo, S.~Pfenninger, Policy decision support
  for renewables deployment through spatially explicit practically optimal
  alternatives, Joule 4~(10) (2020) 2185--2207.
\newblock \href {https://doi.org/https://doi.org/10.1016/j.joule.2020.08.002}
  {\path{doi:https://doi.org/10.1016/j.joule.2020.08.002}}.

\bibitem{neumann2021}
F.~Neumann, T.~Brown, The near-optimal feasible space of a renewable power
  system model, Electric Power Systems Research 190 (2021).
\newblock \href {https://doi.org/https://doi.org/10.1016/j.epsr.2020.106690}
  {\path{doi:https://doi.org/10.1016/j.epsr.2020.106690}}.

\bibitem{tyndp_2018}
Ento-E, \href{https://tyndp.entsoe.eu/tyndp2018/}{Ten year network development
  plan} (2018).
\newline\urlprefix\url{https://tyndp.entsoe.eu/tyndp2018/}

\bibitem{EEA_emissions}
\href{https://www.eea.europa.eu/ims/greenhouse-gas-emission-intensity-of-1}{Greenhouse
  gas emission intensity of electricity generation in europe} (2021).
\newline\urlprefix\url{https://www.eea.europa.eu/ims/greenhouse-gas-emission-intensity-of-1}

\bibitem{schlott2021carbon}
M.~Schlott, O.~E. Sayed, M.~Bilousova, F.~Hofmann, A.~Kies, H.~St{\"o}cker,
  Carbon leakage in a european power system with inhomogeneous carbon prices,
  arXiv (2021).
\newblock \href {https://doi.org/https://arxiv.org/abs/2105.05669}
  {\path{doi:https://arxiv.org/abs/2105.05669}}.

\bibitem{policy101}
R.~G. Newell,
  \href{https://www.rff.org/publications/explainers/federal-climate-policy-101/}{Federal
  climate policy 101: Reducing emissions} (2021).
\newline\urlprefix\url{https://www.rff.org/publications/explainers/federal-climate-policy-101/}

\bibitem{eu_ets}
Eu emissions trading system (eu ets) (2021).

\bibitem{horsch2018pypsa}
J.~H{\"o}rsch, F.~Hofmann, D.~Schlachtberger, T.~Brown, Pypsa-eur: An open
  optimisation model of the european transmission system, Energy Strategy
  Reviews 22 (2018) 207--215.
\newblock \href {https://doi.org/https://doi.org/10.1016/j.esr.2018.08.012}
  {\path{doi:https://doi.org/10.1016/j.esr.2018.08.012}}.

\bibitem{powerplantmatching}
F.~Hofmann, J.~Hörsch, Fresna - powerplantmatching v.0.4.1 (2019).
\newblock \href {https://doi.org/https://doi.org/10.5281/zenodo.3358985}
  {\path{doi:https://doi.org/10.5281/zenodo.3358985}}.

\bibitem{IRENA}
\href{https://www.irena.org/Statistics/Download-Data}{Anual statistics} (2019).
\newline\urlprefix\url{https://www.irena.org/Statistics/Download-Data}

\bibitem{Data_catalogue}
\href{https://ens.dk/en/our-services/projections-and-models/technology-data}{Technology
  data catalogues} (2019).
\newline\urlprefix\url{https://ens.dk/en/our-services/projections-and-models/technology-data}

\bibitem{ENTSOE}
ENSTO-E, Data portal,
  \url{"https://www.entsoe.eu/data/data-portal/consumption/"} (2020).

\bibitem{solonen2012efficient}
A.~Solonen, P.~Ollinaho, M.~Laine, H.~Haario, J.~Tamminen, H.~J{\"a}rvinen,
  Efficient mcmc for climate model parameter estimation: Parallel adaptive
  chains and early rejection, Bayesian Analysis 7~(3) (2012) 715--736.
\newblock \href {https://doi.org/https://doi.org/10.1214/12-BA724}
  {\path{doi:https://doi.org/10.1214/12-BA724}}.

\end{thebibliography}

\clearpage
\onecolumn

\appendix

\section{Unused CO$_2$ emissions}\label{app:unusedCO2}

\begin{figure}[htb]
               \includegraphics[width=1\textwidth,trim={1.5cm 0.6cm 2.5cm 1.2cm},clip] {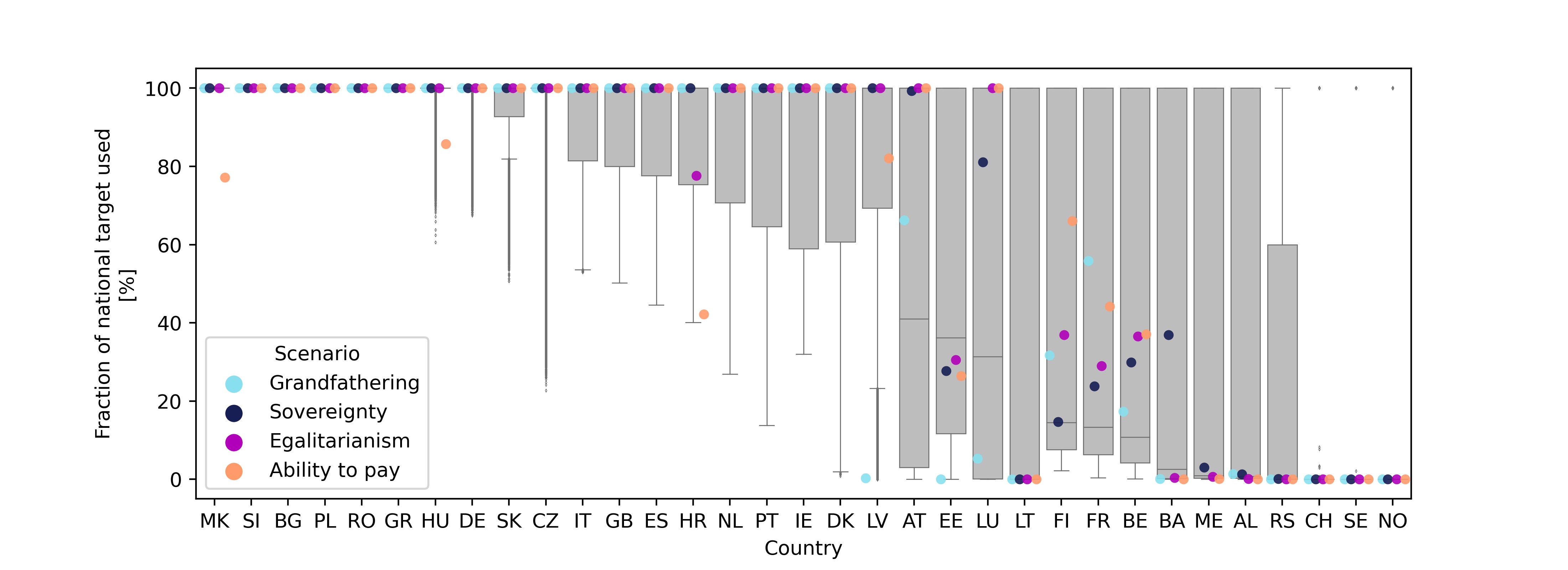}
               \includegraphics[width=1\textwidth,trim={0cm 0cm 0cm 0cm},clip]{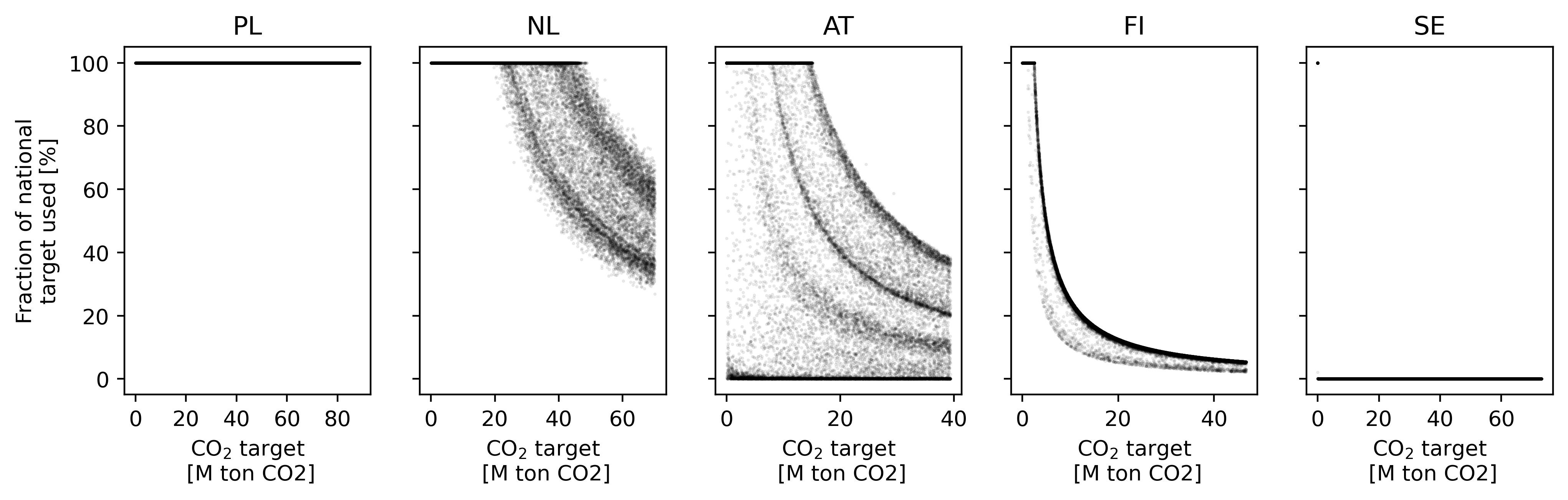}
               \caption{\textbf{Utilization of national emission targets -} a) A box plot of the utilization of the national reduction targets for the individual countries across all scenarios. A value of 100\% means that the country is emitting as much CO$_2$ as their reduction target allows, whereas 0\% indicates that the country has no emissions although the CO$_2$ reduction target is not 0. The reduction configurations Efficiency 55\% and Efficiency 70\% are not included on the figure as they per definition utilize all assigned emissions. 
               b) Example countries. Reduction target utilization plotted against the total amount of target emissions.}
               \label{fig:unused_co2}
\end{figure}

\begin{figure}[htb]
               \includegraphics[width=0.5\textwidth,trim={3.2cm 3cm 1.1cm 2cm} ,clip] {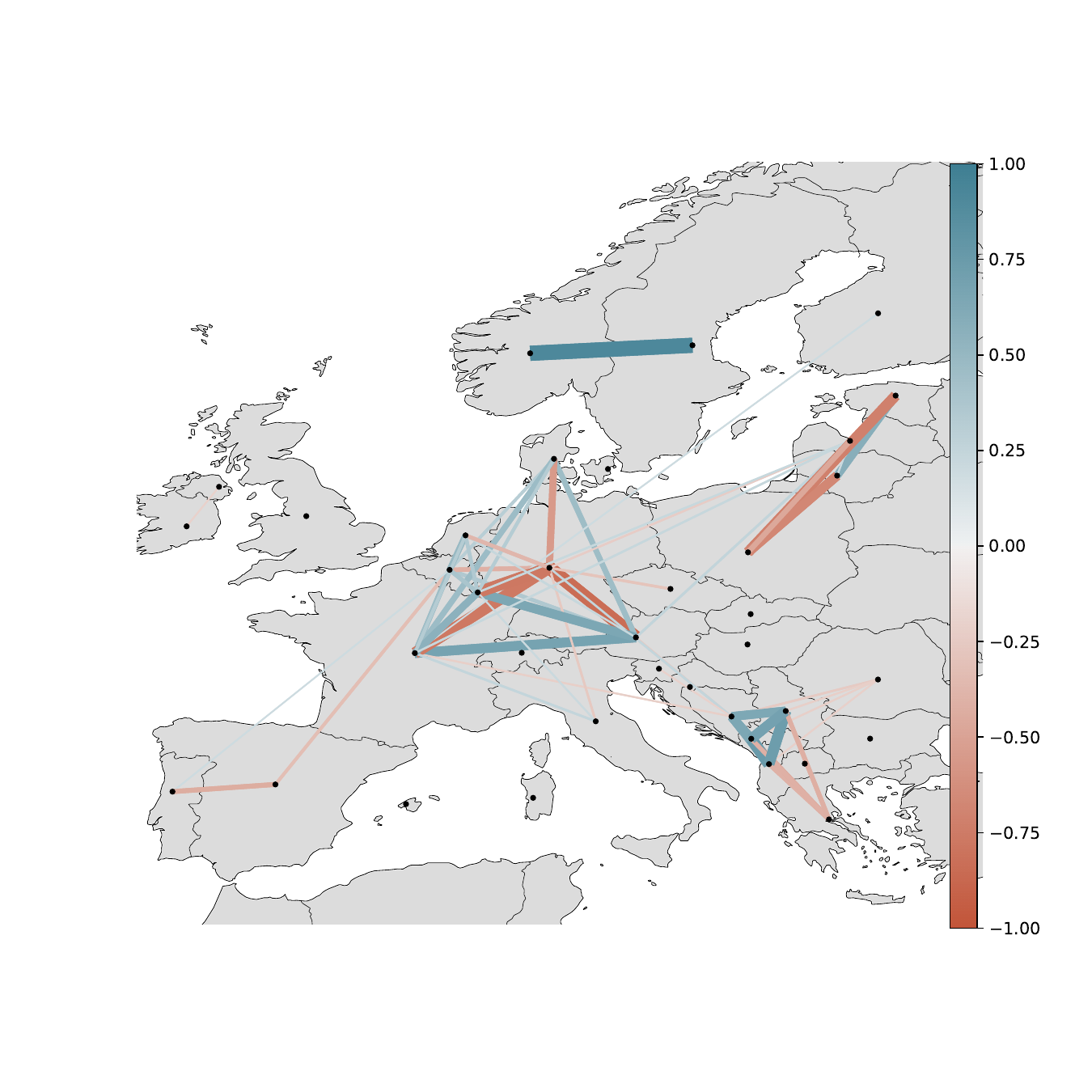}
               \includegraphics[width=0.5\textwidth,trim={1cm 1.5cm 1.5cm 1.5cm} ,clip]{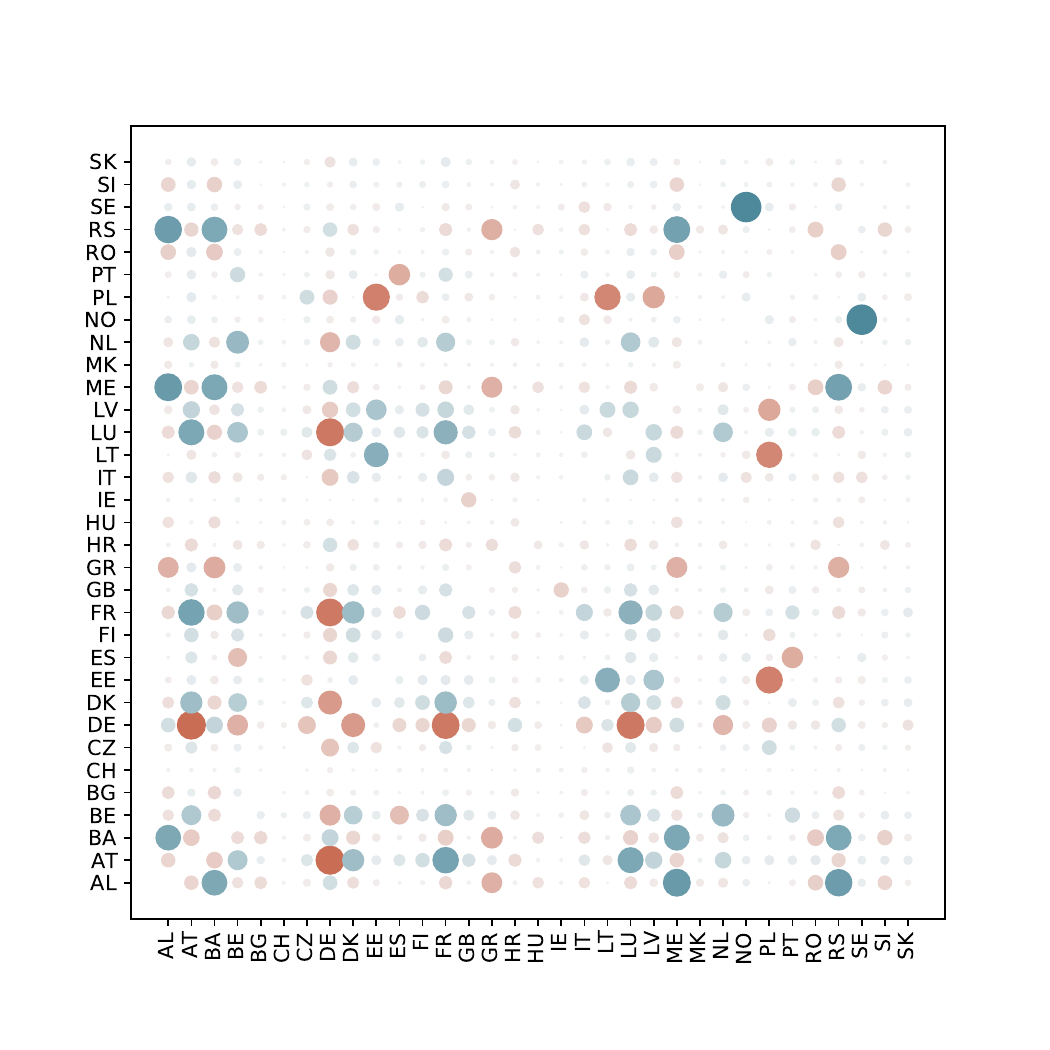}
               \caption{\textbf{Correlation of emissions -} The data used to create this figure is the Pearson correlation of the national CO$_2$ emission across all samples. a) CO$_2$ emission correlations shown on a map of the model countries. Correlation strength and direction is indicated by the link color and size. Correlations below 0.2 has been removed for clarity. b) A matrix plot of CO$_2$ emission correlation for all model countries. The strong positive correlation between Sweden and Norway’s emissions is an artifact, as these two countries have zero emissions with almost all approaches and configurations.}
               \label{fig:co2_correlation}
\end{figure}

In this study, national emission targets were assigned to countries giving the countries the option to either use all allowable emissions or simply leave them unused if it is economically optimal. The top panel of Figure \ref{fig:unused_co2} shows a box plot of the utilization of the national reduction targets for the individual countries across all configurations of national reduction target allocations. Here, a value of 100\% means that the country is emitting as much CO$_2$ as their reduction target allows, whereas 0\% indicates that the country has no emissions although the CO$_2$ reduction target is not 0. Below the box plot, the probability of full reduction target utilization has been shown for five illustrative countries (Figure \ref{fig:unused_co2}). It features three distinct behavioral patterns for the countries. The countries either a) emit as much as the national target allows no matter what, b) sometimes overperform on the target or c) never have any emission. These different patterns are clearly highlighted in the lower panels: Poland is seen always emitting as much as the national target allows. The Netherlands, Austria, and Finland can be seen to be frequently over-performing, with less than the allowable emissions. The curved lines appearing in these figures correspond to the countries capping out at a certain emission level. Finland can be seen having a clear upper bound to how much CO$_2$ they will choose to emit, whereas Austria appears to have a wide range of possible outcomes available within the criteria used. This behavior is very likely a result of the strong correlation between Austria’s emissions with the emission from Germany and France found in Figure \ref{fig:co2_correlation}. In configurations where Germany is using less than their assigned emissions, Austria is providing dispatchable power and therefore itself has higher emissions. In scenarios where Germany and its neighbors have high allowable emissions, the demand for dispatchable power imports from Austria drops, and it becomes cost-optimal for Austria not to use all the assigned emissions.

\clearpage
\section{Model assumptions}\label{app:assumptions}

\begin{table}[htb]
\caption{Technology data. Emissions are given as t CO$_2$ per MWh electricity produced.}
\label{tab:tech_data}
\begin{tabularx}{.9\textwidth}{XXX}
\hline
Technology      & Efficiency    &   Emissions       \\ 
                &   \%          &   ton CO$_2$/MWhe    \\  \hline       
OCGT            & 41            &   0.49            \\ 
CCGT            & 58            &   0.34            \\
Coal plant            & 33            &   1.00            \\
Lignite plant         & 33            &   1.24            \\
Oil plant             & 35            &   0.77            \\
Electrolysis    & 66            &   0               \\
Fuel Cell       & 50            &   0               \\
Battery inverter & 96           &   0               \\ \hline  

\end{tabularx}
\end{table}

\begin{table}[htbp]
\caption{Technology costs of new technologies.}
\label{tab:tech_cost}
\begin{tabularx}{1\textwidth}{Xrrrr}
\hline
Technology                              & Capital cost  & FOM       &   VOM     & Lifetime   \\ 
                                        &  Eur/kW       & \%/year   &  Eur/MWh  & years \\ \hline
OCGT                                    & 435.2         & 1.78      & 4.5       & 25  \\
Offshore wind turbine                   &  1573.2       & 2.29      & 2.67      & 30    \\
Offshore wind AC connection submarine   &  2685.0*      & 0         &  0        & 30    \\
Offshore wind AC connection underground &  1342.0*      & 0         &  0        & 30    \\
Offshore wind AC station                &  250.0        & 0         & 0         & 30    \\
Offshore wind DC connection submarine   &  2000.0*      & 0         & 0         & 30    \\
Offshore wind DC connection underground &  1000.0*      & 0         & 0         & 30    \\
Offshore wind DC station                &  400          & 0         & 0         & 30    \\
Onshore wind                            & 1035.6        & 1.22      & 1.35      & 30    \\ 
Utility scale solar PV                  & 376.3         & 1.93      & 0         & 40   \\
Electrolysis                            & 550.0         & 5.0       & 0         & 25  \\
Fuel Cell                               & 1100.0        & 5.0       & 0         & 10  \\
Hydrogen storage tank                   & 44.0**        & 1.11      & 0         & 30 \\
Hydrogen underground storage            & 2.0**         & 0         & 0         & 100 \\
Battery inverter                        & 160.0         & 0.34      & 0         & 25 \\
Battery storage                         & 142.0**       & 0         & 0         & 25 \\ \hline
* Eur/MW/km \\
** Eur/kWh \\
\end{tabularx}
\end{table}

\begin{table}[htbp]
\caption{Existing generator technology capacities by 2030 in MW}
\label{tab:brownfield_cap}
\begin{tabularx}{1.1\textwidth}{lXXXXXXXXXX}
 &  Offshore wind &   Onshore wind & Run off river & Solar PV &     CCGT &     OCGT &     Coal &  Lignite &   Nuclear &      Oil \\ \hline   
 AT      &      0.0 &  3132.7 & 4478.5 &  1438.6 &  2481.7 & 1313.5 &   991.5 &      0.0 &      0.0 &    0.0 \\
BA      &      0.0 &    50.6 &    0.0 &     0.0 &     0.0 &    0.0 &     0.0 &      0.0 &      0.0 &    0.0 \\
BE      &   1185.9 &  2074.8 &   59.0 &  3984.5 &  3801.9 & 1460.6 &  1524.8 &      0.0 &   5925.8 &    0.0 \\
BG      &      0.0 &   691.0 &   22.4 &  1029.0 &     0.0 &  782.0 &  4963.7 &   3993.0 &   2000.0 &    0.0 \\
CH      &      0.0 &    63.0 & 5280.0 &  2171.0 &     0.0 &    0.0 &     0.0 &      0.0 &   3430.0 &    0.0 \\
CZ      &      0.0 &   316.2 &   40.2 &  2074.3 &   336.8 &    0.0 &  7184.7 &    725.7 &   2660.0 &    0.0 \\
DE      &   6396.0 & 52447.0 & 2997.0 & 45179.0 & 18120.9 & 8044.3 & 28069.4 &  20833.5 &  15788.4 & 3696.4 \\
DK      &   1708.1 &  4431.2 &    0.0 &   991.0 &   100.0 & 1427.4 &  3629.9 &      0.0 &      0.0 &  665.0 \\
EE      &      0.0 &   329.8 &    0.0 &    25.4 &   173.0 &  250.0 &     0.0 &      0.0 &      0.0 & 2111.0 \\
ES      &      0.0 & 23433.1 &   16.4 &  4753.5 & 24344.3 & 2942.6 &  6519.7 &   3081.2 &   7572.6 & 3533.4 \\
FI      &     67.0 &  1971.3 & 1289.6 &   123.0 &   648.0 &  677.7 &  3039.7 &      0.0 &   2784.0 & 1225.4 \\
FR      &      0.0 & 14898.1 & 5780.8 &  9604.0 &  5611.0 & 1066.0 &  4293.3 &      0.0 &  63130.0 & 7172.1 \\
GB      &   8212.7 & 13553.9 &  685.2 & 13107.3 & 32824.3 &  921.5 & 14475.0 &      0.0 &  11261.0 & 2801.9 \\
GR      &      0.0 &  2877.5 &  103.1 &  2650.6 &  4482.0 &  417.0 &  1550.0 &   3905.0 &      0.0 &    0.0 \\
HR      &      0.0 &   580.3 &  278.7 &    67.4 &   369.6 &   82.5 &   304.3 &      0.0 &      0.0 &  647.8 \\
HU      &      0.0 &   335.0 &   19.7 &   724.0 &  1259.2 & 2368.7 &    42.3 &   1180.2 &   1886.8 &  410.0 \\
IE      &     25.2 &  3650.9 &  216.0 &    21.8 &  2946.0 & 1320.0 &   855.0 &      0.0 &      0.0 &  907.0 \\
IT      &      0.0 & 10230.2 & 6563.7 & 20073.6 & 34438.1 & 6491.8 & 10926.5 &      0.0 &      0.0 & 6145.0 \\
LT      &      0.0 &   532.0 &    0.0 &    81.9 &     0.0 & 1575.0 &     0.0 &      0.0 &      0.0 &    0.0 \\
LU      &      0.0 &   114.2 &   30.9 &   124.7 &   350.5 &    0.0 &     0.0 &      0.0 &      0.0 &    0.0 \\
LV      &      0.0 &    62.9 &  642.1 &     0.0 &  1025.0 &    0.0 &     0.0 &      0.0 &      0.0 &    0.0 \\
ME      &      0.0 &   118.0 &    0.0 &     0.0 &     0.0 &    0.0 &     0.0 &      0.0 &      0.0 &    0.0 \\
MK      &      0.0 &    37.0 &   41.6 &    17.0 &     0.0 &    0.0 &     0.0 &    824.0 &      0.0 &    0.0 \\
NL      &    957.0 &  3491.0 &    0.0 &  4522.0 & 13582.0 & 3991.0 &  5591.0 &      0.0 &    492.0 &    0.0 \\
NO      &      0.0 &  1708.0 &    0.0 &    53.4 &   450.0 &  773.1 &     0.0 &      0.0 &      0.0 &    0.0 \\
PL      &      0.0 &  5762.1 &   14.4 &   562.0 &   326.0 & 1032.9 & 21588.5 &   9406.0 &      0.0 &  345.0 \\
PT      &      0.0 &  5172.4 & 1615.5 &   665.4 &  3829.0 &    0.0 &  1756.0 &      0.0 &      0.0 &    0.0 \\
RO      &      0.0 &  3243.0 &  870.4 &  1385.7 &  1080.0 & 2282.0 &  1506.0 &   4779.2 &   1298.0 &   87.5 \\
RS      &      0.0 &    25.0 &    0.0 &     0.0 &     0.0 &    0.0 &     0.0 &      0.0 &      0.0 &    0.0 \\
SE      &    204.0 &  7097.0 & 1955.9 &   481.0 &   708.0 &    0.0 &   130.0 &      0.0 &   9532.0 & 2135.0 \\
SI      &      0.0 &     0.0 &  861.3 &   251.8 &   832.0 &  449.0 &   246.0 &    944.0 &    727.0 &  143.6 \\
SK      &      0.0 &     0.0 &  641.3 &   533.0 &   648.0 &    0.0 &   440.0 &    486.0 &   1940.0 &    0.0 \\ \hline   
\end{tabularx}
\end{table}

\clearpage
\section{Scenarios with equal realized emissions}\label{sec:all55scenarios}

\begin{figure}[!htb]
   \includegraphics[width=0.45\textwidth,trim={0cm 0cm 0cm 0cm},clip]{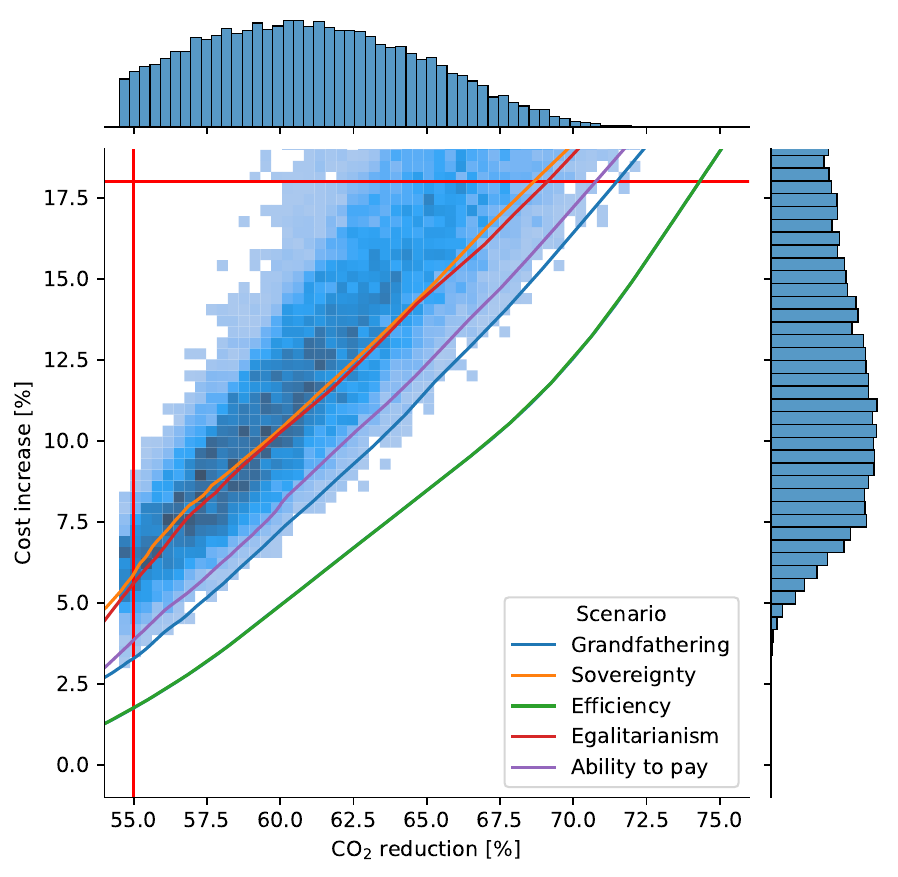}
   \includegraphics[width=0.5\textwidth,trim={0cm 0cm 0cm 0cm},clip]{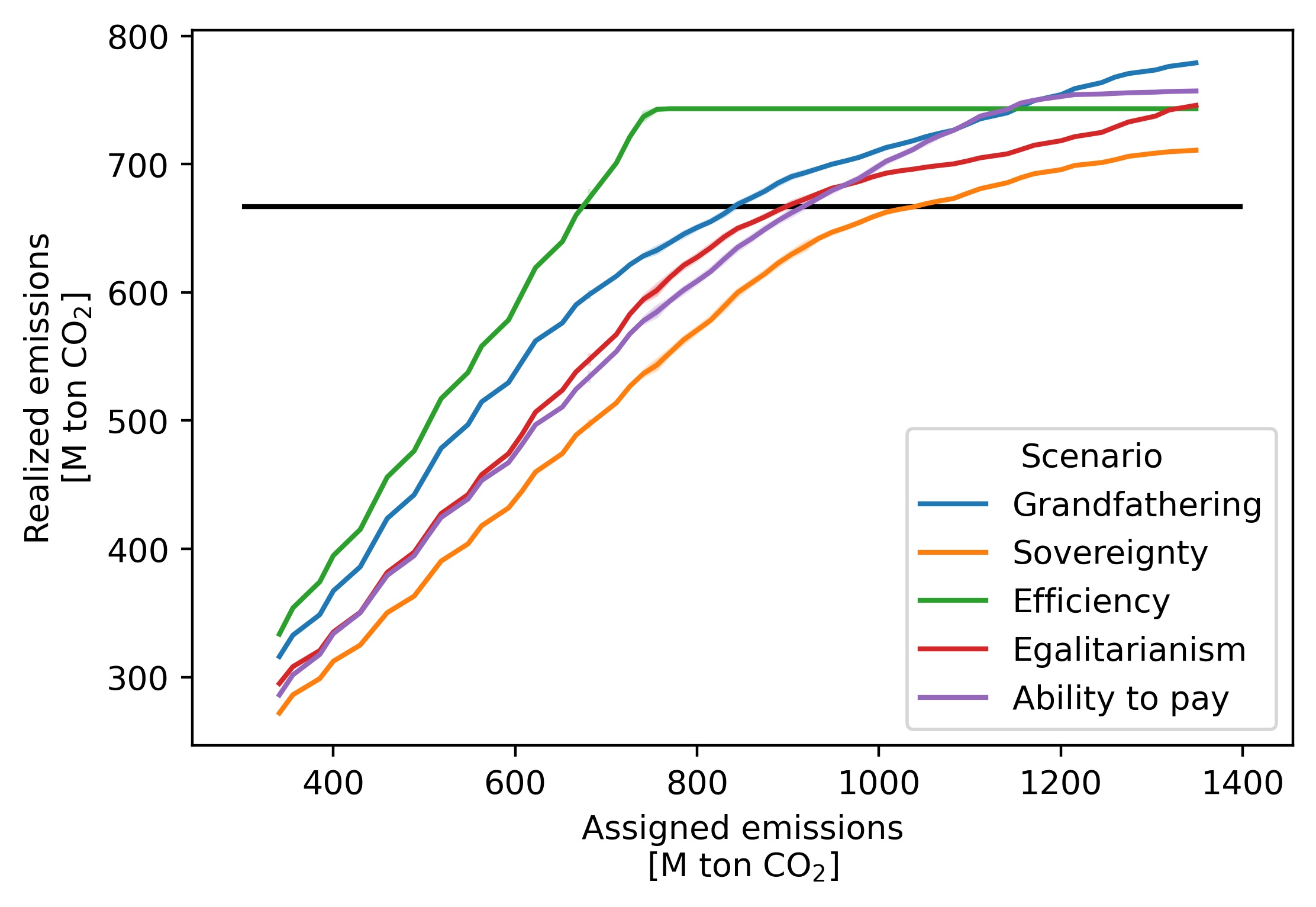}
   \caption{a) Shows the relationship between cost increase and CO$_2$ reduction for all configuration strategies. b) Realized CO$_2$ emissions from all configuration strategies plotted against the sum of assigned CO$_2$ targets. The horizontal black line represents the CO$_2$ budget associated with a 55\% CO$_2$ reduction.}
   \label{fig:all55_scenarios}
\end{figure}

In this work, a decision was made to compare scenarios (see Table \ref{tab:scenarios}) that assign CO$_2$ targets summing up to the same total CO$_2$ budget. The CO$_2$ targets are not always fully utilised as some countries find it economically favorable to reduce emissions despite being assigned allowable emissions. The result is that the scenarios compared have different global emission reductions although having the same CO$_2$ budget as seen in Figure \ref{fig:co2red_cost}. The shares of unused CO$_2$ emissions is seen in Figure \ref{fig:unused_co2}.
Alternatively, one could design the scenarios to have equal realized global emissions. By adjusting the CO$_2$ budget for the individual scenarios, realized emissions corresponding to a 55\% emission reduction can be achieved. Figure \ref{fig:all55_scenarios} b) shows the relationship between assigned emissions and realized emissions for all configuration strategies listed in Table \ref{tab:scenarios}. The figure clearly shows configuration strategies other than Efficiency having lower realized than assigned emissions.

\begin{figure}[!htbp]
   \includegraphics[width=1\textwidth,trim={1.5cm 0.6cm 2.5cm 1.2cm},clip]{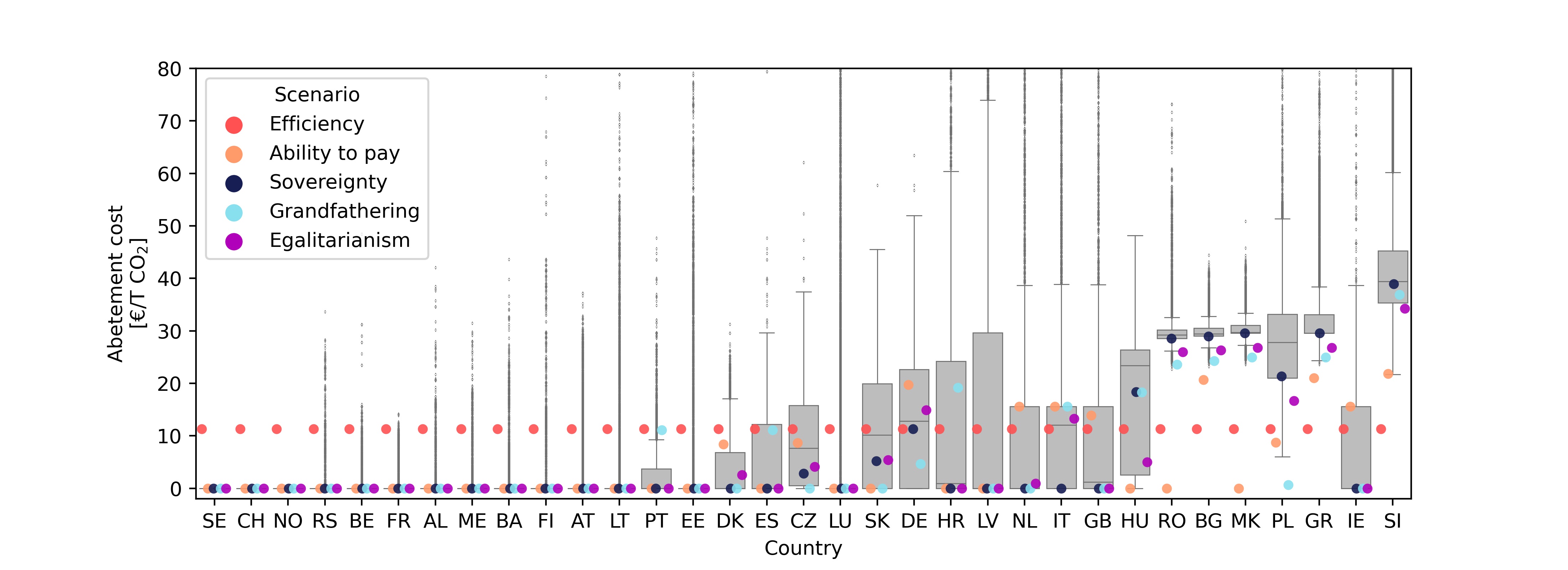}
   \includegraphics[width=1\textwidth,trim={1.5cm 0.6cm 2.5cm 1.2cm},clip]{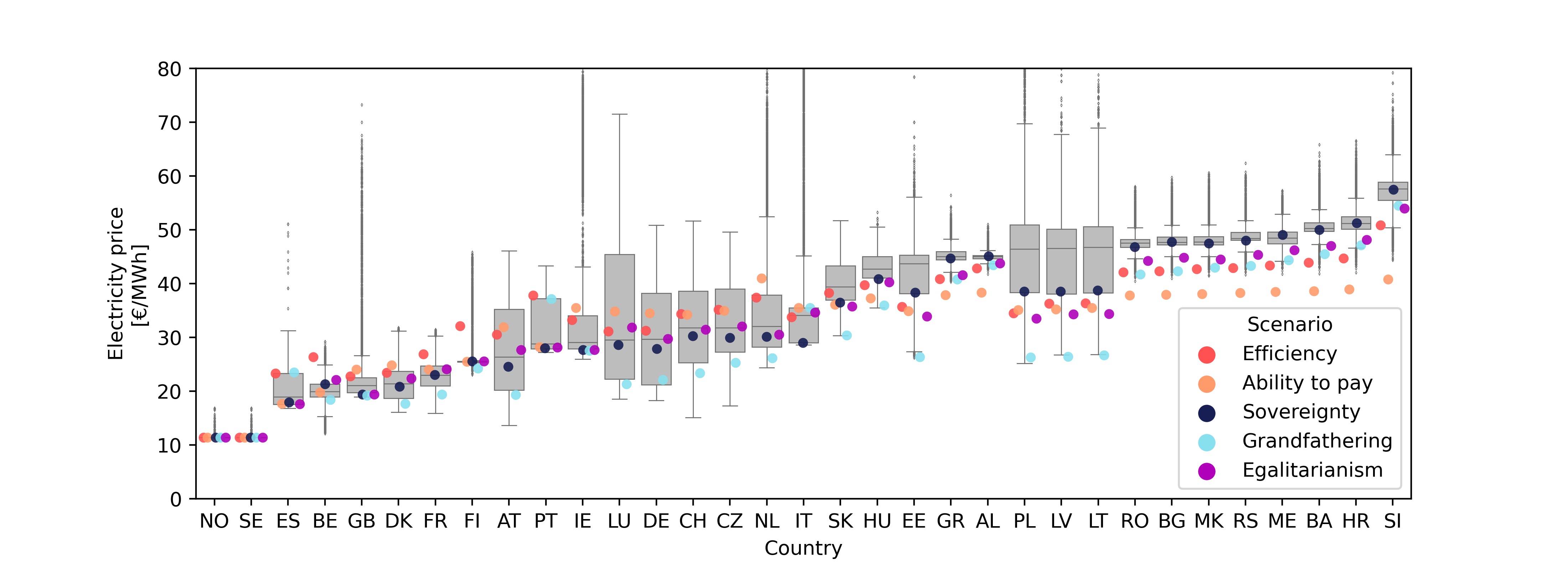}
   \caption{The three figures shows results for scenarios that all have realized emissions corresponding to a 55\% CO$_2$ reduction. Panel a) shows relative emissions on country level. Panel b) shows abatement cost, and panel c) shows electricity prices.}
   \label{fig:all55_boxplots}
\end{figure}

In this section the result found by selecting scenarios with realized emissions corresponding to a 55\% CO$_2$ reduction is shown. Figure \ref{fig:all55_boxplots} shows figures corresponding to Figure \ref{fig:co2_mwh_box} and \ref{fig:prices}, but the configuration strategies shown all have realized emissions corresponding to a 55\% CO$_2$ reduction. Thus the configurations shown here all have higher realized emissions then the configurations used in Figure \ref{fig:co2_mwh_box} and \ref{fig:prices}. The configurations studied in this section with realized emissions corresponding to a 55\% CO$_2$ reduction will be referred to as the "55\% realized" configurations, whereas the original scenarios with equal CO$_2$ budgets will be referred to as the "55\% CO$_2$ budget" configurations.

\begin{figure}[!htb]
   \includegraphics[width=1\textwidth,trim={1.5cm 0.6cm 2.5cm 1.2cm},clip]{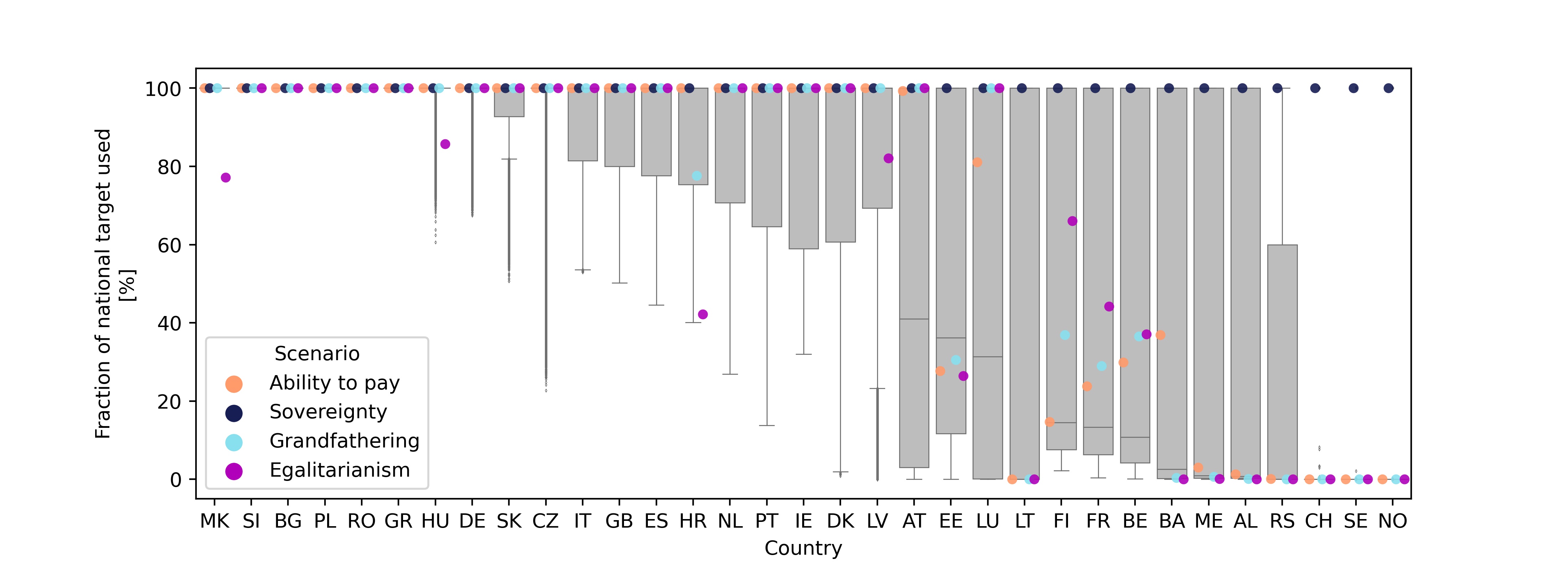}
   \caption{Unused emissions given as the fraction of the national target used, for scenarios that all have realized emissions corresponding to a 55\% CO$_2$ reduction.}
   \label{fig:all55_unused_co2}
\end{figure}

Comparing the "55\% CO$_2$ budget" configurations found on Figures \ref{fig:co2_mwh_box} and  \ref{fig:prices} with the "55\% realized" configurations, found on Figure \ref{fig:all55_boxplots}, the "55\% realized" configurations are found to be more equal. Across all measures shown in Figure \ref{fig:all55_boxplots}, the "55\% realized" configurations deviate less from each other than the "55\% CO$_2$ budget" configurations does on Figures \ref{fig:co2_mwh_box} and  \ref{fig:prices}. This behavior is expected as the "55\% realized" configurations redistribute unused CO$_2$ from countries finding it favorable to reduce emissions, to countries where higher emissions are desired. Thus, the "55\% realized" configurations become more similar to the Efficiency configuration compared to the "55\% CO$_2$ budget" configurations. This is also reflected in the significant cost decrease between the "55\% realized" and "55\% CO$_2$ budget" configuration shown on Figure \ref{fig:all55_scenarios} a). Furthermore, Figure \ref{fig:all55_unused_co2} shows how the "55\% realized" scenarios have much less unused CO$_2$ emissions than the "55\% CO$_2$ budget" configuration.

\clearpage
\section{Detailed model results}\label{sec:detailed_results}

\begin{figure}[p]
   \includegraphics[width=1.1\textwidth,trim={0cm 0cm 0cm 0cm},clip]{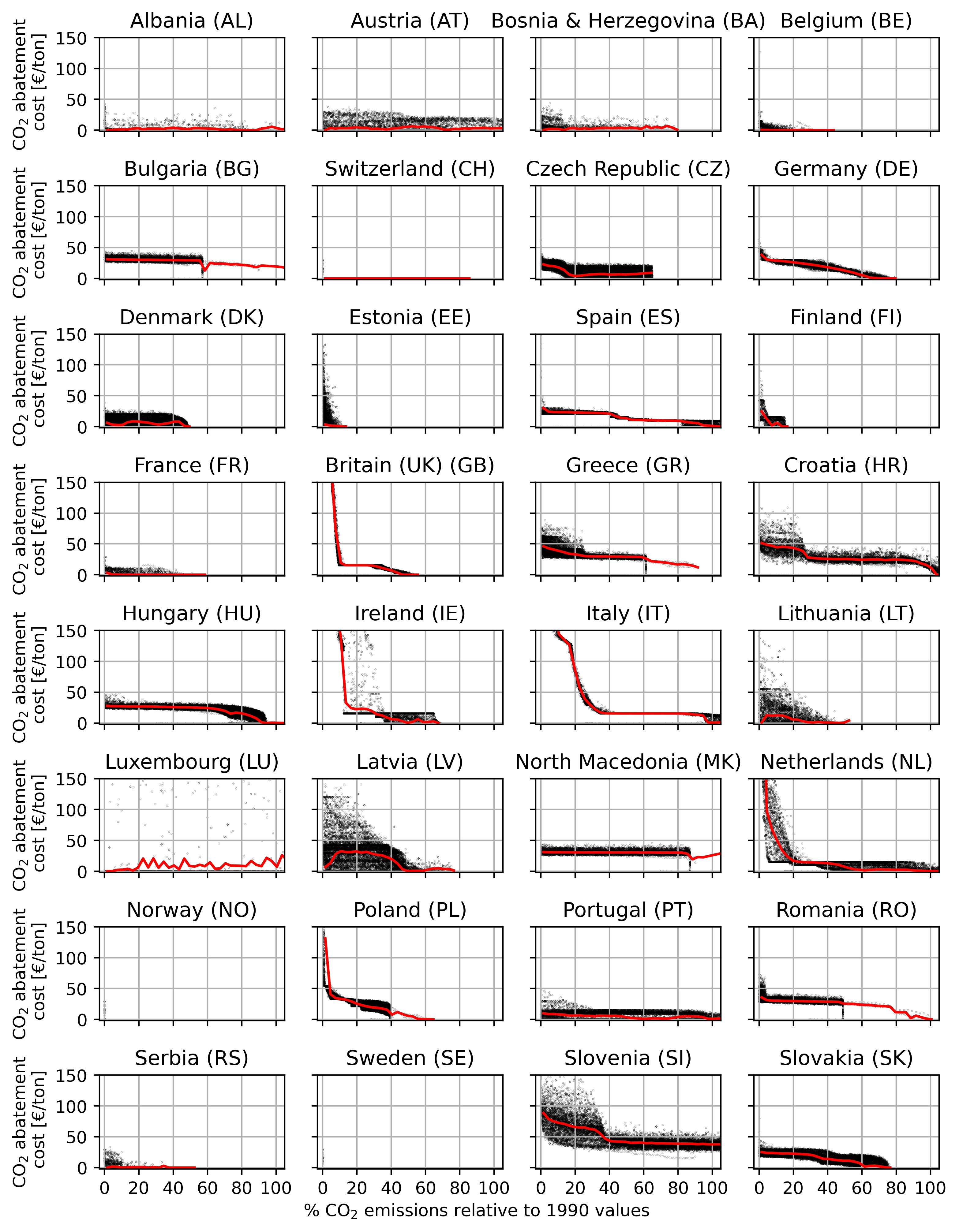}
   \caption{CO$_2$ abatement cost for all model countries plotted against the CO$_2$ reduction level relative to 1990 values. Sample points are shown with black dots and the sample mean is shown with a red line. 1990 emission values not available for Montenegro (ME) and the country has therefore been excluded from the figure. }
   \label{fig:all_abatement_price}
\end{figure}

\begin{figure}[p]
   \includegraphics[width=1.1\textwidth,trim={0cm 0cm 0cm 0cm},clip]{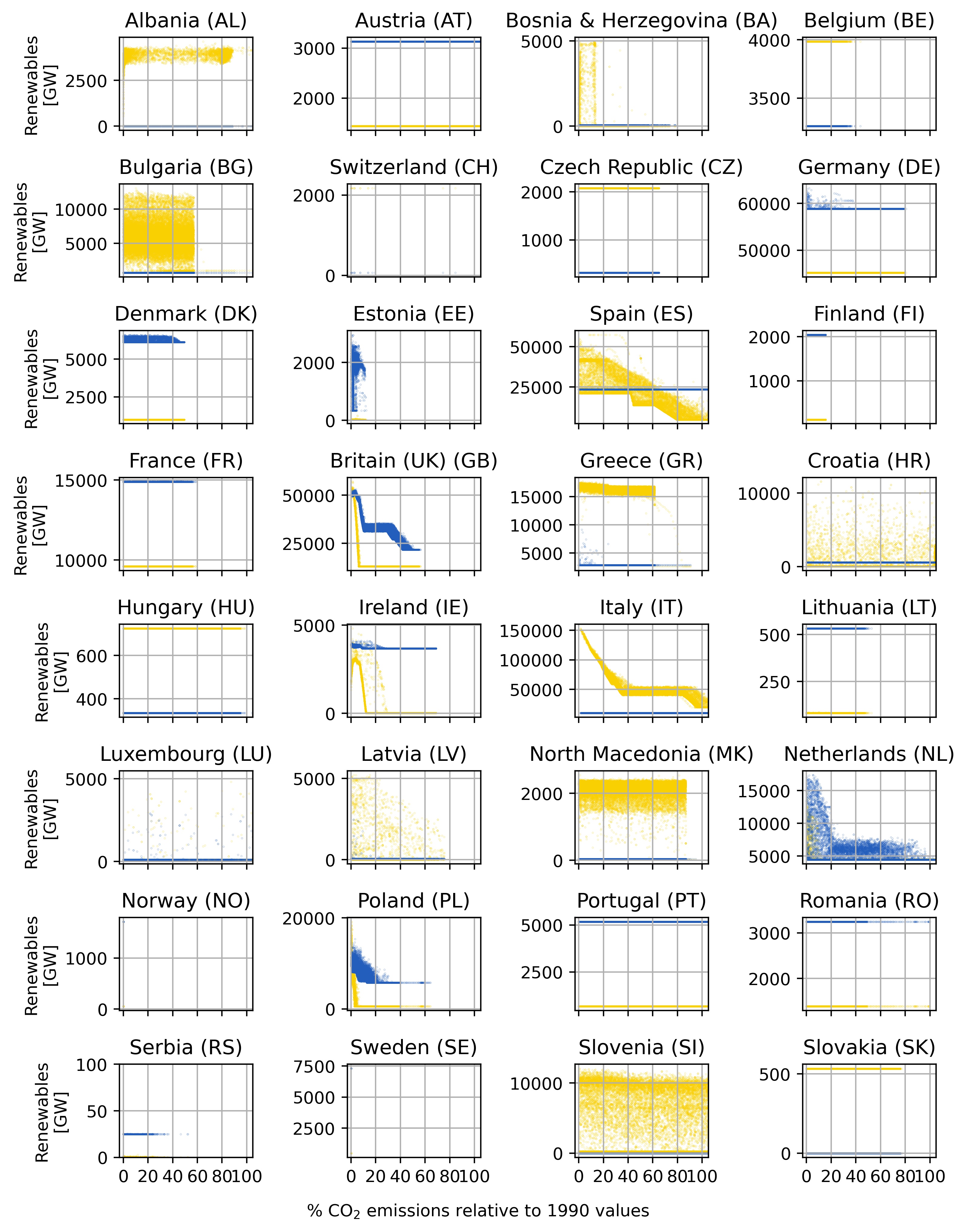}
   \caption{Renewable energy generator capacities plotted against CO$_2$ reduction levels. Every sample is shown as a single dot. Solar PV capacity is indicated by yellow dots, and wind turbine capacity with blue dots. 1990 emission values not available for Montenegro (ME) and the country has therefore been excluded from the figure. }
   \label{fig:all_renewable_cap}
\end{figure}

\begin{figure}[p]
   \includegraphics[width=1.1\textwidth,trim={0cm 0cm 0cm 0cm},clip]{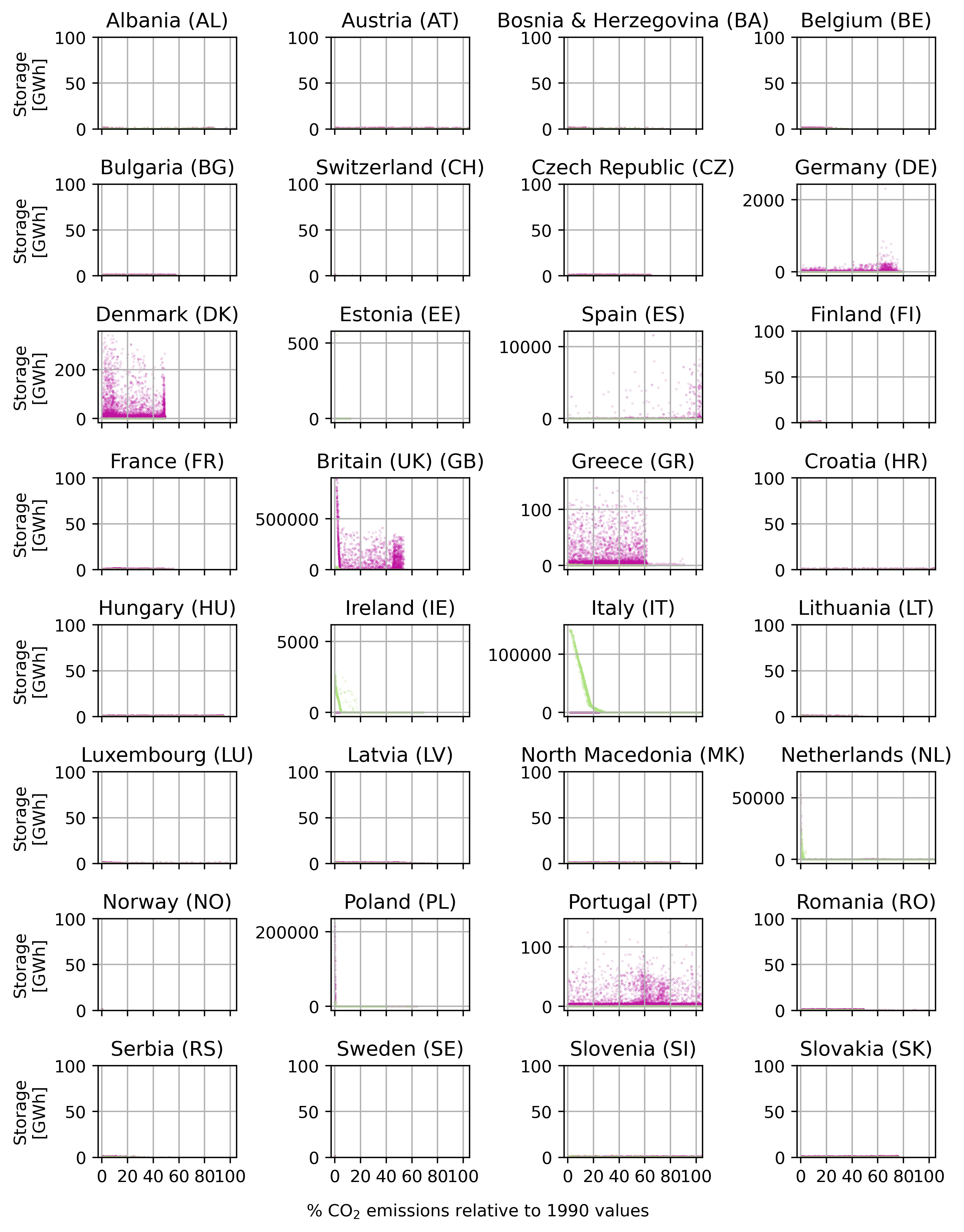}
   \caption{Storage capacity for all model countries measured in GWh storage capacity. Every sample is shown as a single dot. Battery storage is indicated by green dots, while H2 storage is shown with purple. 1990 emission values not available for Montenegro (ME) and the country has therefore been excluded from the figure. }
   \label{fig:all_storage_cap}
\end{figure}

\begin{figure}[p]
   \includegraphics[width=1.1\textwidth,trim={0cm 0cm 0cm 0cm},clip]{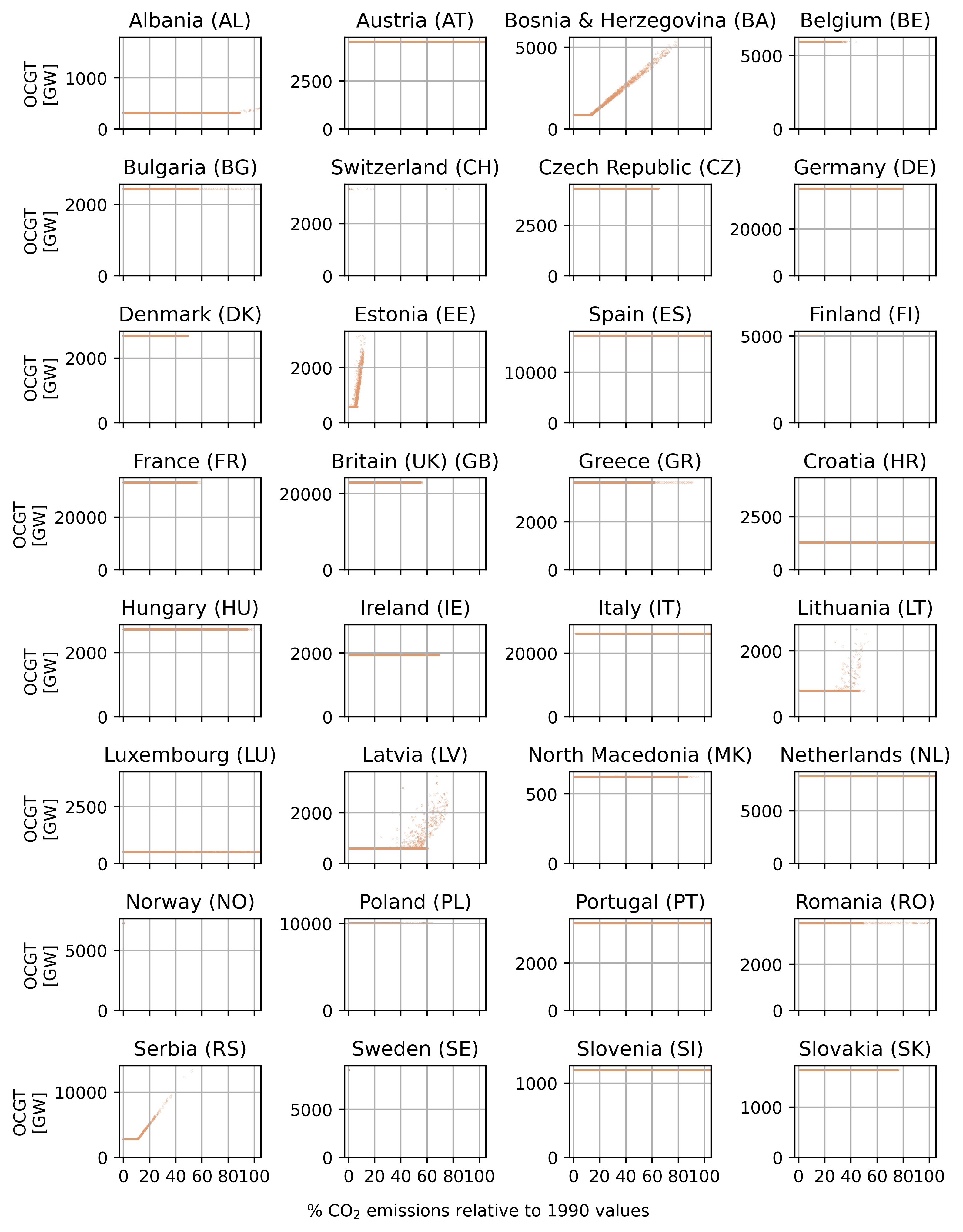}
   \caption{Open cycle gas turbine (OCGT) capacity for all model countries. Every sample is shown as a single dot. OCGT is the only extendable non-renewable energy source included in the model. 1990 emission values not available for Montenegro (ME) and the country has therefore been excluded from the figure.}
   \label{fig:all_OCGT_cap}
\end{figure}

\begin{figure}[!htb]
   \includegraphics[width=1\textwidth,trim={1cm 0cm 2.5cm 0cm},clip]{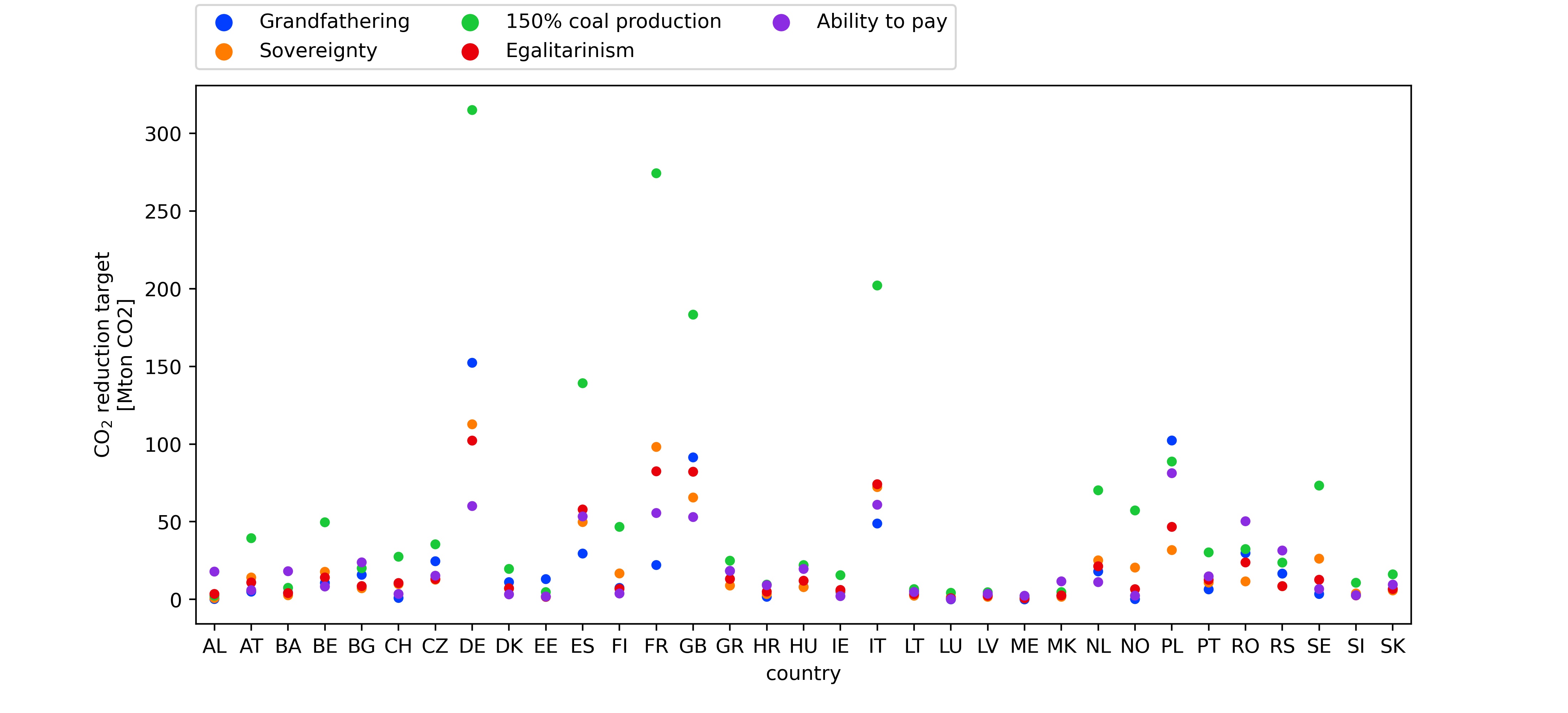}
   \caption{CO2 target layouts for the scenarios used and the 150\% coal production upper limit. The 150\% coal upper limit is calculated as the load of each nation multiplied by an emission factor of 0.45[tCO$_2$ per MWh] times 1.5.  }
   \label{fig:co2_scenarios}
\end{figure}

               
\end{document}